\documentclass[a4paper]{article}

\usepackage{indentfirst}
\usepackage{fullpage}
\usepackage{amsmath}
\usepackage{amsthm}
\usepackage{latexsym}
\usepackage{epsf}
\usepackage{graphicx, color}

\makeatletter
\renewcommand\paragraph{%
  \@startsection{paragraph}{4}{0mm}%
   {-\baselineskip}%
   {.5\baselineskip}%
   {\normalfont\normalsize\bfseries}}
\makeatother

\date{}

\begin{document}

\title{\Large \bf \boldmath\ An oscillating motion of a red blood cell and a neutrally
buoyant particle in Poiseuille flow in a narrow channel}

\author{
Lingling Shi, Yao Yu, Tsorng-Whay Pan\footnotemark[1], Roland Glowinski\\
%\footnotemark[]
Department of Mathematics, University of Houston, Houston, TX 77204, USA}

\renewcommand{\thefootnote}{\fnsymbol{footnote}}
%\footnotetext[1]{lingling@math.uh.edu}
\footnotetext[1]{Corresponding author. E-mail address: pan@math.uh.edu}
%\footnotetext[3]{roland@math.uh.edu}
%\footnotetext[2]{Project supported by an NSF grant DMS-9973318.}
\newtheorem{remark}{Remark}[section]

\newcommand{\bff}{{\bf f}}
\newcommand{\bn}{{\bf n}}
\newcommand{\bnabla}{{\boldsymbol{\nabla}}}
\newcommand{\bg}{{\bf g}}
\newcommand{\bu}{{\bf u}}
\newcommand{\bD}{{\bf D}}
\newcommand{\bF}{{\bf F}}
\newcommand{\bG}{{\bf G}}
\newcommand{\bU}{{\bf U}}
\newcommand{\bv}{{\bf v}}
\newcommand{\bx}{{\bf x}}
\newcommand{\bV}{{\bf V}}
\newcommand{\bz}{{\bf 0}}
\newcommand{\cth}{{\mathcal{T}_h}}
\newcommand{\calt}{{\mathcal{T}}}
\newcommand{\g}{{\bf g}}
\newcommand{\gx}{{\Gamma_{-} (0, \Delta t)}}
\newcommand{\Gx}{{{\overrightarrow{Gx}}^{\perp}}}
\newcommand{\dsum}{{\displaystyle\sum}}
\newcommand{\into}{{\displaystyle{\int_{\Omega}}}}
\newcommand{\intG}{{\displaystyle{\int_{\Gamma}}}}
\newcommand{\oo}{{\overline{\Omega}}}
\newcommand{\ox}{{\Omega \times (0, \Delta t)}}
\newcommand{\oxot}{{\Omega \times (0,T)}}
\newcommand{\obo}{{\Omega \backslash \overline{B(0)}}}
\newcommand{\obt}{{\Omega \backslash \overline{B(t)}}}
\newcommand{\R}{{I\!\!R}}
\newcommand{\blambda}{{\boldsymbol{\lambda}}}
\newcommand{\bmu}{{\boldsymbol{\mu}}}
\newcommand{\bsigma}{{\boldsymbol{\sigma}}}

\maketitle

\begin{abstract}
Two motions of oscillation and vacillating breathing (swing) of a red blood cell have been observed 
in bounded Poiseuille flows (Phys. Rev. E {\bf 85}, 16307 (2012)). 
To understand such motions, we have studied the oscillating motion of a neutrally buoyant 
rigid particle of the same shape in Poiseuille flow in a narrow channel and obtained 
that the crucial point is to have the particle interacting 
with Poiseuille flow with its mass center moving up and down in the channel central region. 
Since the mass center of the cell migrates toward the channel central region, its oscillating motion of the
inclination angle is similar to the aforementioned motion as long as the cell keeps the shape 
of long body. But as the up-and-down oscillation of the cell mass center damps out, the oscillating motion 
of the inclination angle also damps out and the cell inclination angle approaches to a fixed angle.   

\end{abstract}

\baselineskip 14pt

\setlength{\parindent}{1.5em}

\setcounter{section}{0}

\vskip 2ex

\section{Introduction}
The microhydrodynamics of deformable entities such as lipid vesicles  
and red blood cells in flows has received
increasing attention experimentally, theoretically, and numerically
in recent years.  
Lipid vesicles (e.g., see \cite{Hass1997}-\cite{Kim2012}) 
%\cite{Hass1997,Kantsler2005,Seifert1999,Misbah2006,Kraus1996,Beaucourt2004,Noguchi2004,Noguchi2005,Noguchi2007,Danker2007,Kim2012}
and red blood cells (e.g., see \cite{Fischer1978}-\cite{Abkarian2007})
%\cite{Fischer1978,Keller1982,Tran-Son-Tay1984,Pozrikidis2003,Skotheim2007,Abkarian2007}
show phenomenologically similar behaviors in shear flows, which are (i) a tank treading rotation 
with a stationary shape and a finite inclination angle with respect to the 
flow direction, (ii) an unsteady tumbling motion, and (iii) vacillating breathing
(the long axis undergoes oscillation  about the flow, while the shape shows breathing).
In \cite{Shi2012a,Shi2012b}, two motions of oscillation and vacillating breathing of a red blood cell  
in bounded Poiseuille flows have been observed in low Reynolds number regime. 
The vacillating breathing motion is actually a combination of oscillation and deformation 
of the cell. The motion of oscillation of a blood cell
in bounded Poiseuille flows is more interesting. In \cite{Sugihara-Seki1993}, the 
motion of an elliptical cylinder particle immersed in an incompressible Newtonian fluid in a narrow channel 
has been examined numerically in Stokes flow regime under the assumption that no external forces
or torques act on the elliptical cylinder. Two interesting motions of a particle of elliptic 
shape are that it either tumbles (i.e., keep rotating) while the mass center always stays
away from the centerline or rotates changing itself direction as the mass 
center crosses the centerline (oscillating motion). Even though the oscillating motion of a particle of elliptic 
shape is a periodic motion in Stokes flow regime, it closely resembles to the one associated 
with a red blood cell in \cite{Shi2012a,Shi2012b}. 

In this paper we have compared the motions of a neutrally buoyant particle and a red blood cell in 
Poiseuille flow in the low Reynolds number regime to understand the oscillating motions 
in \cite{Shi2012a,Shi2012b} and to find out the difference between the cell motion and the particle 
motion in a narrow channel. For the motion of a neutrally buoyant particle of either biconcave or 
elliptical shape, we have obtained similar oscillating motion at the channel central region in 
Poiseuille flow when the particle mass center is placed at the centerline initially; but such motion 
is a transition since, later on, the particle migrates away from the centerline and starts tumbling 
as expected. Indeed in bounded Poiseuille flows (e.g., see \cite{Segre1961}-\cite{Chen2012}), 
%\cite{Segre1961,Segre1962,Asmolov1999,Asmolov2002,Eloot2004,Yang2005,Chen2012})
a neutrally buoyant particle migrates laterally to its equilibrium position
between the channel centerline and the wall due to the competition between the shear 
gradient of the Poiseuille flow and the wall effect.  For a  neutrally buoyant cylinder 
of elliptical shape, it tumbles while migrating toward and then staying at its 
equilibrium position (e.g., see \cite{Chen2012}). Concerning the motion of a red blood cell, 
after the cell migrates to the central region the oscillating motion occurs due to the interaction 
between the cell and the profile of the Poiseuille flow just like an elliptic shape particle
in \cite{Sugihara-Seki1993} if the cell can maintain a long body shape.  Specifically during the 
oscillating motion for the both cases, both mass centers move up and down in the channel central region. 
But unlike the periodic up-and-down motion of the elliptical cylinder mass center in Stokes flow 
in \cite{Sugihara-Seki1993}, the cell oscillating motion damps out while its mass center 
approaches to an equilibrium height at the central region in a narrow channel  and the cell inclination angle
approaches to a fixed angle. For the cases of a neutrally buoyant cylinder of either biconcave shape or 
elliptical shape  in the low Reynolds number regime, its mass center moves up and down 
about the centerline with an increasing amplitude and then moves away from 
the central region and toward its equilibrium position between the channel centerline and the wall.  
Thus the up and down motion of the mass center in the channel central region triggers the oscillation motion
of a long body entity in Poiseuille flows.

The scheme of this paper is as follows: We discuss the models and numerical methods briefly
in Section 2. In Section 3, we have investigated the effects of the channel height, the bending
rigidity of the membrane, the initial position on the motion of a red blood cell and a neutrally
buoyant particle in Poiseuille flow in a narrow channel. The conclusions are summarized in Section 4.

\section{Models and methods}
\subsection{Model and method for a red blood cell}

\begin{figure}[h]
\begin{center}
\leavevmode \epsfxsize=5 true in \epsffile{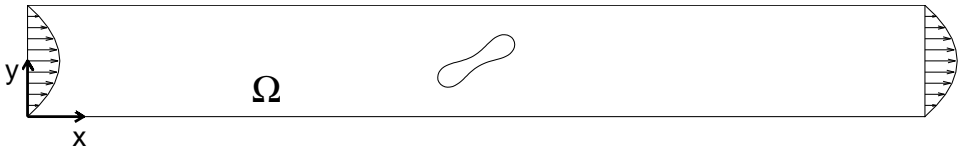}
\end{center}
\caption{Schematic of a single red blood cell in plane Poiseuille flow with the computational domain 
$\Omega$.} \label{fig.1}
\end{figure}

An RBC with the viscosity of the cytoplasm same as that of the blood plasma is suspended in a fluid
domain $\Omega$ filled with blood plasma which is incompressible and Newtonian as in Figure \ref{fig.1}. 
For some $T>0$, the
governing equations for the fluid-cell system are the Navier-Stokes equations

\begin{eqnarray}
&&\rho_f \left( \dfrac{\partial \bu}{\partial t} + \bu \cdot\nabla \bu \right) =-\nabla p+\mu \triangle \bu+\bff,
\text{    in   }  \Omega \times (0, T),  \label{eqn:1a}  \\
&& \nabla \cdot \bu=0,\text{    in   }  \Omega \times (0, T). \label{eqn:1b}
\end{eqnarray}

\noindent Equations (\ref{eqn:1a}) and (\ref{eqn:1b}) are completed by the following boundary and
initial conditions:
\begin{eqnarray}
&& \bu={\bf 0} \text{ on the top and bottom of $\Omega$ and $\bu$ is periodic in the $x$ direction,} \label{eqn:1c}  \\
&& \bu(\bx,0)=\bu_0(\bx) ,  \text{    in   }  \Omega\label{eqn:1d}
\end{eqnarray}
\noindent where $\bu$ and $p$ are the fluid velocity and pressure, respectively,  
$\rho_f$ is the fluid density, and $\mu$ is the fluid viscosity, which is assumed to be
constant for the entire fluid. In eq. (\ref{eqn:1a}), $\bff$ is a body force which is the sum of $\bff_p$
and $\bff_B$ where $\bff_p$ is the pressure gradient pointing in the $x$ direction and $\bff_B$ accounts
for the force acting on the interface between fluid and cell. In eq. (\ref{eqn:1d}), $\bu_0(\bx)$ is the initial fluid velocity.
 
\begin{figure}
\begin{center}
\leavevmode \epsfxsize=2 true in \epsffile{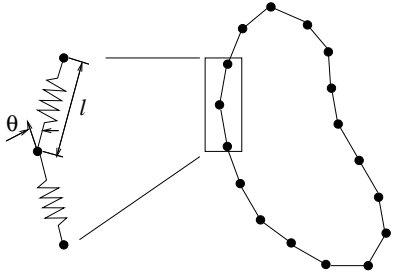}
\end{center}
\caption{The elastic spring model of the RBC membrane.} \label{fig.2}
\end{figure}

The deformability and the elasticity of the RBC are due to the skeleton architecture of the membrane.
A two-dimensional elastic spring model used in \cite{Tsubota2006} is considered in this paper to describe
the deformable behavior of the RBCs. Based on this model, the RBC membrane can be viewed as membrane
particles connecting with the neighboring membrane particles by springs, as shown in Figure \ref{fig.2}.
Energy stores in the spring due to the change of the length $l$ of the spring with respect to its
reference length $l_0$ and the change in angle $\theta$ between two neighboring springs. The total
energy of the RBC membrane, $E=E_l+E_b$, is the sum of the total energy for stretch and compression
and the total energy for the bending which, in particular, are

\begin{equation}
E_{l}=\frac{k_{l}}{2}\sum_{i=1}^{N}(\frac{l_{i}-l_{0}}{l_{0}})^{2}, \ \  E_{b}=\frac{k_{b}}{2}\sum_{i=1}^{N}\tan^{2}(\theta_{i}/2). \label{eqn:2}
\end{equation}
In equation (\ref{eqn:2}), $N$ is the total number of
the spring elements, and $k_{l}$ and $k_{b}$ are spring constants
for changes in length and bending angle, respectively.
The cell shape is stimulated by reducing the total area of the circle of radius $R_0=2.8 \rm{\ \mu m}$
through a penalty function
\begin{equation}
\Gamma_{s}=\frac{k_{s}}{2}(\frac{s-s_{e}}{s_{e}})^{2}\label{eqn:2c}
\end{equation}
where $s$ and $s_{e}$ are the time dependent area of the RBC and the specified area of the RBC, respectively, 
and the total energy is modified as $E+\Gamma_{s}$.
Based on the principle of virtual work the force acting on the
$i$th membrane particle now is
\begin{equation}
{\bf F}_{i}=-\frac{\partial(E+\Gamma_{s})}{\partial\label{eqn:2d}
{\bf r}_{i}}
\end{equation}
where ${\bf r}_{i}$ is the position of the $i$th membrane
particle. The value of the swelling ratio of an RBC in this paper is
defined by $s^{*}=s_e/(\pi R_0^2)$.

The motion of the RBCs in the fluid flow is simulated by combining the immersed boundary 
method \cite{Peskin1977, Peskin1980, Peskin2002} and the aforementioned elastic spring model 
for RBC membrane.  The Navier-Stokes equations for fluid flow have been solved by using an operator
splitting technique and finite element method \cite{Glowinskibook,Shi2010IJNMF}
with a regular triangular mesh so that the specialized fast solver, such as FISHPAK by
Adams et al. \cite{Adams}, can be used to solve 
the fluid flow. The methodologies have been validated in previous 
studies \cite{Shi2012a,Shi2010IJNMF}.

\subsection{Model and method for a neutrally buoyant particle}

We suppose that $\Omega$ is filled with an incompressible viscous Newtonian fluid of the 
density $\rho_f$ and viscosity $\mu$. Let $B(t)$ be a freely moving rigid neutrally 
buoyant particle in a fluid as in Figure \ref{fig.3}. 
The boundaries of $\Omega$ and $B(t)$ are denoted by $\Gamma$ and $\partial B$, respectively.
For some $T>0,$ the governing equations for the fluid-particle system are the Navier-Stokes
equations for the fluid flow:
\begin{figure}[t]
\begin{center}
\leavevmode \epsfxsize=4.5 true in \epsffile{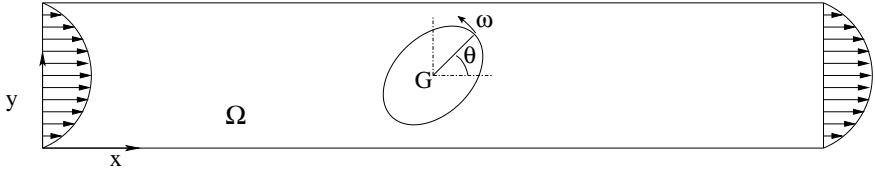}
\end{center}
\caption{Schematic of a neutrally buoyant particle in plane Poiseuille flow with the domain $\Omega$.} \label{fig.3}
\end{figure}
\begin{eqnarray}
&& \rho_{f}( \dfrac{\partial \bu }{\partial t} +
( \bu \cdot \bnabla )\bu ) 
=\bff -\bnabla p +2\mu \bnabla \cdot \bD(\bu)
   \ \  in \ \Omega \backslash \overline{B(t)},\, t\in(0,T), \label{eqn:2.1.1}\\
&& \nabla \cdot \textbf{u}=0 \ \  in \ \Omega \backslash \overline{B(t)},\, t\in(0,T),\label{eqn:2.1.2}\\
&& \bu(\bx,0)=\bu_{0}(\bx), \ \  \forall \bx \in \Omega \backslash \overline{B(0)}, \ 
with \, \nabla \cdot \bu_{0}=0, \label{eqn:2.1.3}\\
&& \bu={\bf 0} \ \  on \ \Gamma,  \label{eqn:2.1.4}\\
&& \bu= {\bV}_{p} + \omega \times \stackrel{\longrightarrow}{\bG\bx},
 \ \forall \bx \in  \partial B(t), \,\label{eqn:2.1.5}
\end{eqnarray}
where $\bu$ is the flow velocity, $p$ is the pressure,
$2\bD(\bu)=\bnabla \bu + (\bnabla \bu)^{t}$, and $\bff$ is the pressure gradient 
pointing in the $x$ direction.
In (\ref{eqn:2.1.5}), the no-slip condition on the boundary  of the particle,  
$\bV_{p}$ is the translation velocity,  
$\omega \times \stackrel{\longrightarrow}{\bG\bx}=(-\omega (y-G_{2}), \omega (x - G_{1}))^t$ 
where $\omega$ is the angular velocity, 
$\bG=(G_{1},G_{2})^t$ is the mass center and  $\bx=(x,y)^t$ is a
point on the boundary of the particle.

The motion of the  particles is modeled by Newton's laws:
\begin{eqnarray} 
&& M \ \dfrac{d \bV_{p}}{d t} = M\bg + \bF , \, 
 I \dfrac{d \omega}{d t}= {F}^t,\,
 \dfrac{d \bG}{d t} = \bV_{p},\,
 \dfrac{d \theta}{d t} = \omega,\label{eqn:2.1.9}\\
&& \bG(0) = \bG^0, \bV_{p}(0) = \bV^0_{p}, \omega(0) = {\omega}^0, \theta(0)=\theta^0.\label{eqn:2.1.12}
\end{eqnarray} 
In (\ref{eqn:2.1.9})-(\ref{eqn:2.1.12}), $\bg$  is the gravity, $\theta$ is the inclination angle
between the long axis and the horizontal direction,
$M$ and $I$ are the  the mass and the moment of inertia of the particle, respectively;
$\bF$ and ${F}^t$ denote, respectively, the hydrodynamic force and the related torque
imposed on the particle by the fluid given by
\begin{equation} 
\bF =- \int_{\partial B} \bsigma {\bn}\,ds, \ {F}^t=- \int_{\partial B} {\bG\bx} \times \bsigma {\bn}\,ds, \nonumber
\end{equation}
where $\bsigma$ is the stress tensor, $\bx$ is the generic point on the boundary of the particle, 
$\bn$ is the unit normal vector on the boundary of the particle pointing
to the center of the particle and ${\bf a} \times {\bf b}=a_1 b_2 - a_2 b_1$ for the two dimensional cases.

The method of solution for the above fluid-particle interaction is
a combination of a distributed Lagrange multiplier based fictitious domain method,
the operator splitting methods and finite element methods 
(e.g., \cite{Chen2012,Glowinskibook,Pan2002}). 
The basic idea is to imagine that the fluid fills the entire space
inside as well as outside the particle boundaries. The fluid flow problem is then posed on a larger
domain (the ``fictitious domain''). This larger domain is simpler, allowing a simple regular mesh to
be used, which in turn renders use of specialized fast solution techniques as in the previous section. 
The larger domain is also time independent, so the same mesh can be used for the entire simulation, 
eliminating the need for repeated remeshing and projection.
The fluid inside the particle boundary must exhibit a rigid body motion. This constraint is
enforced by using the distributed Lagrange multiplier, which represents the additional body force per
unit volume needed to maintain the rigid body motion inside the particle boundary, much like the
pressure in incompressible fluid flow, whose gradient is the force required to maintain the 
constraint of incompressibility. The computational method has been validated in our 
previous studies \cite{Pan2002} and \cite{Chen2012} for the motion of neutrally buoyant disks   and elliptical cylinders, respectively, in Poiseuille flow. 
In this paper, we  have applied the method to the cases of neutrally buoyant
cylinder of a long body shape freely moving in bounded Poiseuille flows.

\section{Simulation results and discussions}\label{sec.3}

\subsection{Oscillating motion of a single RBC in a narrower channel}\label{sec.3.1}

\begin{figure}
\begin{center}
\leavevmode
\epsfxsize=3in \epsffile{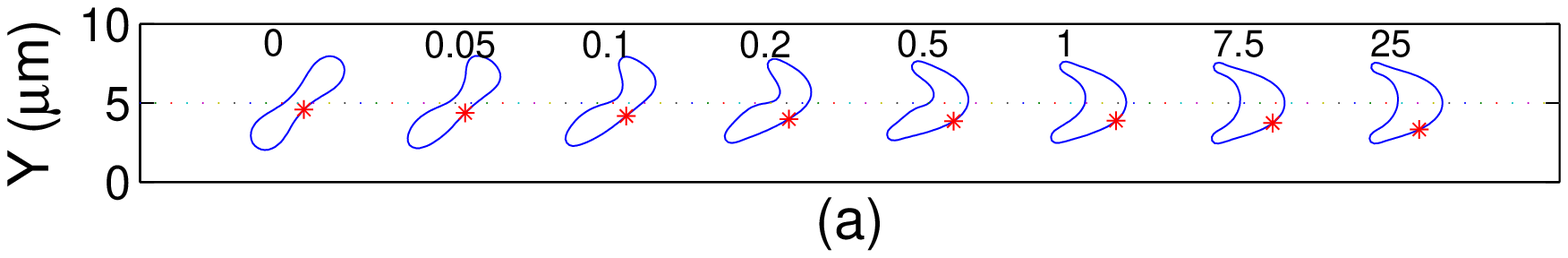}
\epsfxsize=3in \epsffile{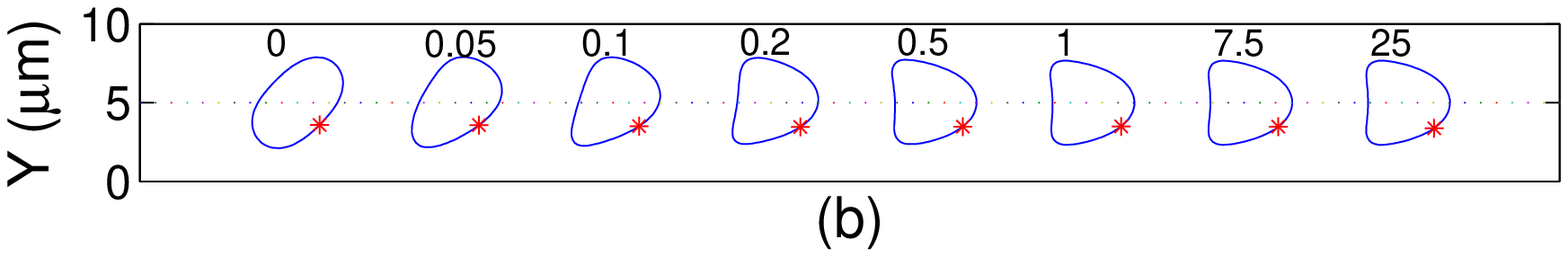}\\
\epsfxsize=3in \epsffile{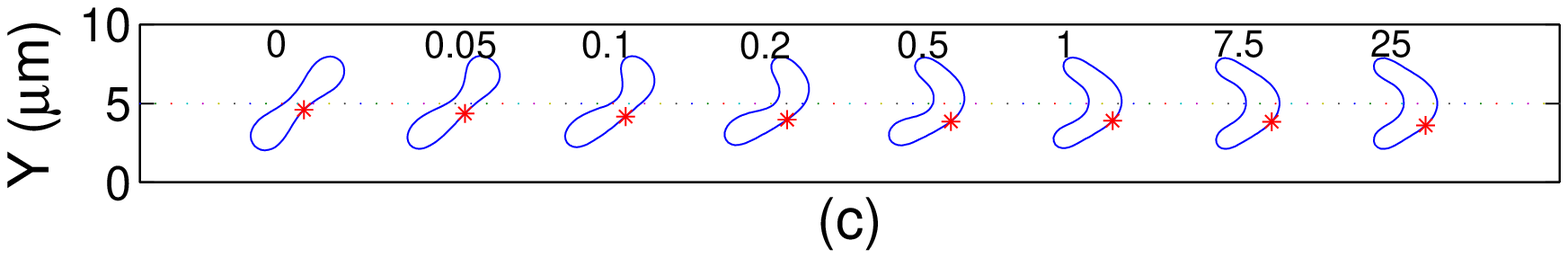}
\epsfxsize=3in \epsffile{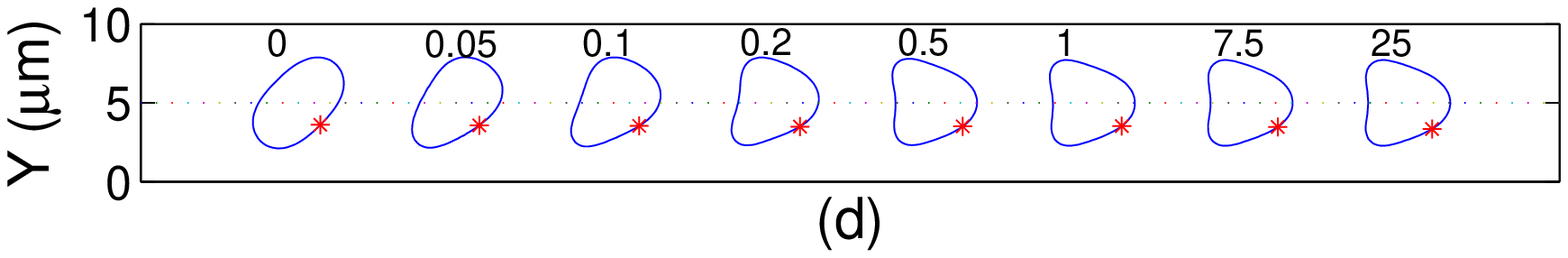}\\
\epsfxsize=3in \epsffile{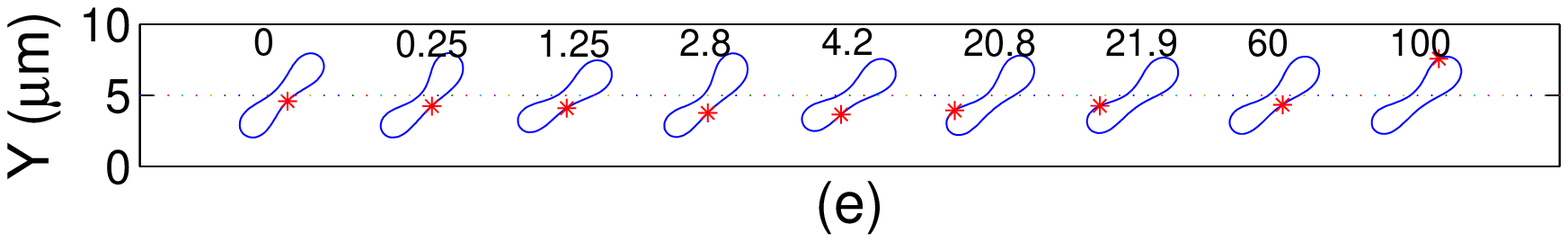}
\epsfxsize=3in \epsffile{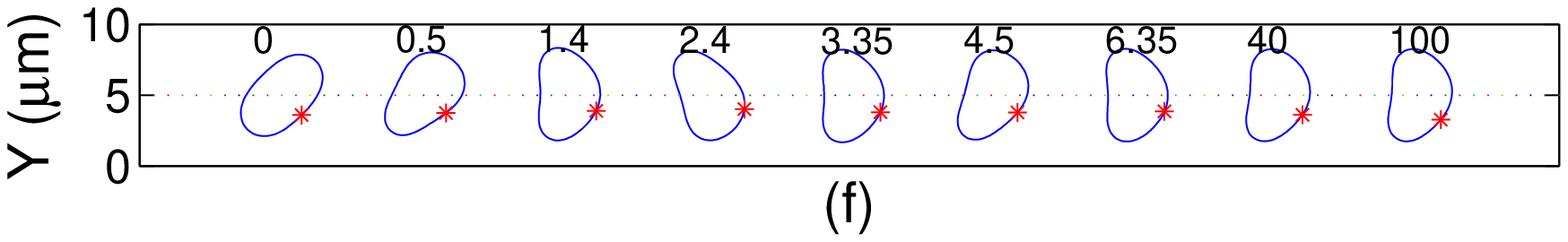}\\
\epsfxsize=3in \epsffile{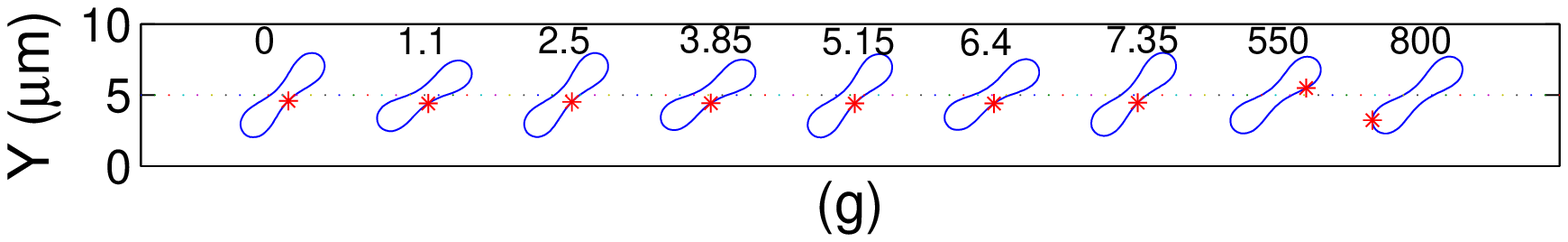}
\epsfxsize=3in \epsffile{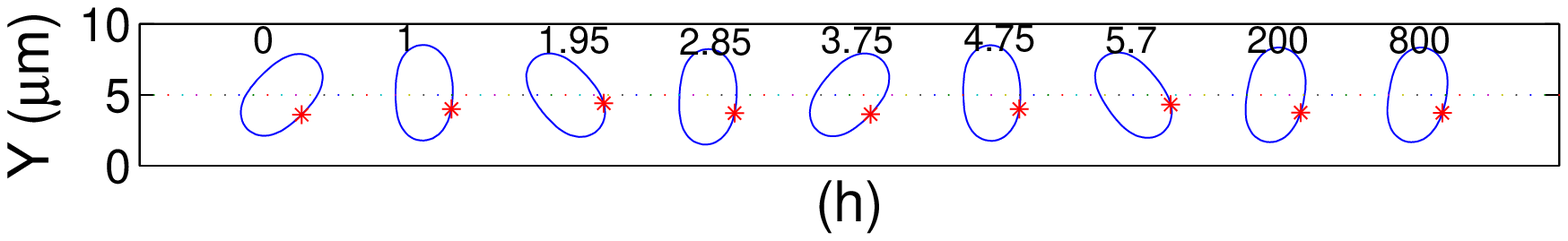}\\
\end{center}
\caption{(Color online). The snapshots of the cell migration in Poiseuille flows for $s^*$=0.481 and 0.9 with different bending
constants at different time ($\rm{ms}$): (a) $s^*$ = 0.481 and 0.1$k_b$, (b) $s^*$ = 0.9 and 0.1$k_b$,
(c) $s^*$ = 0.481 and 1$k_b$, (d) $s^*$ = 0.9 and 1$k_b$, (e) $s^*$ = 0.481 and 10$k_b$, (f) $s^*$ = 0.9 and 10$k_b$,
(g) $s^*$ = 0.481 and 100$k_b$, and (h) $s^*$ = 0.9 and 100$k_b$.
The red asterisk denotes the same node point on the cell membrane. $R_0/w =$ 0.56. 
The initial position is (5,5) at the channel centerline  and the initial inclination angle is $\pi/4$.}
\label{101-umax075-snapshot-45}
%\end{figure}
%\begin{figure}
\begin{center}
\leavevmode
\epsfxsize=2.5in \epsffile{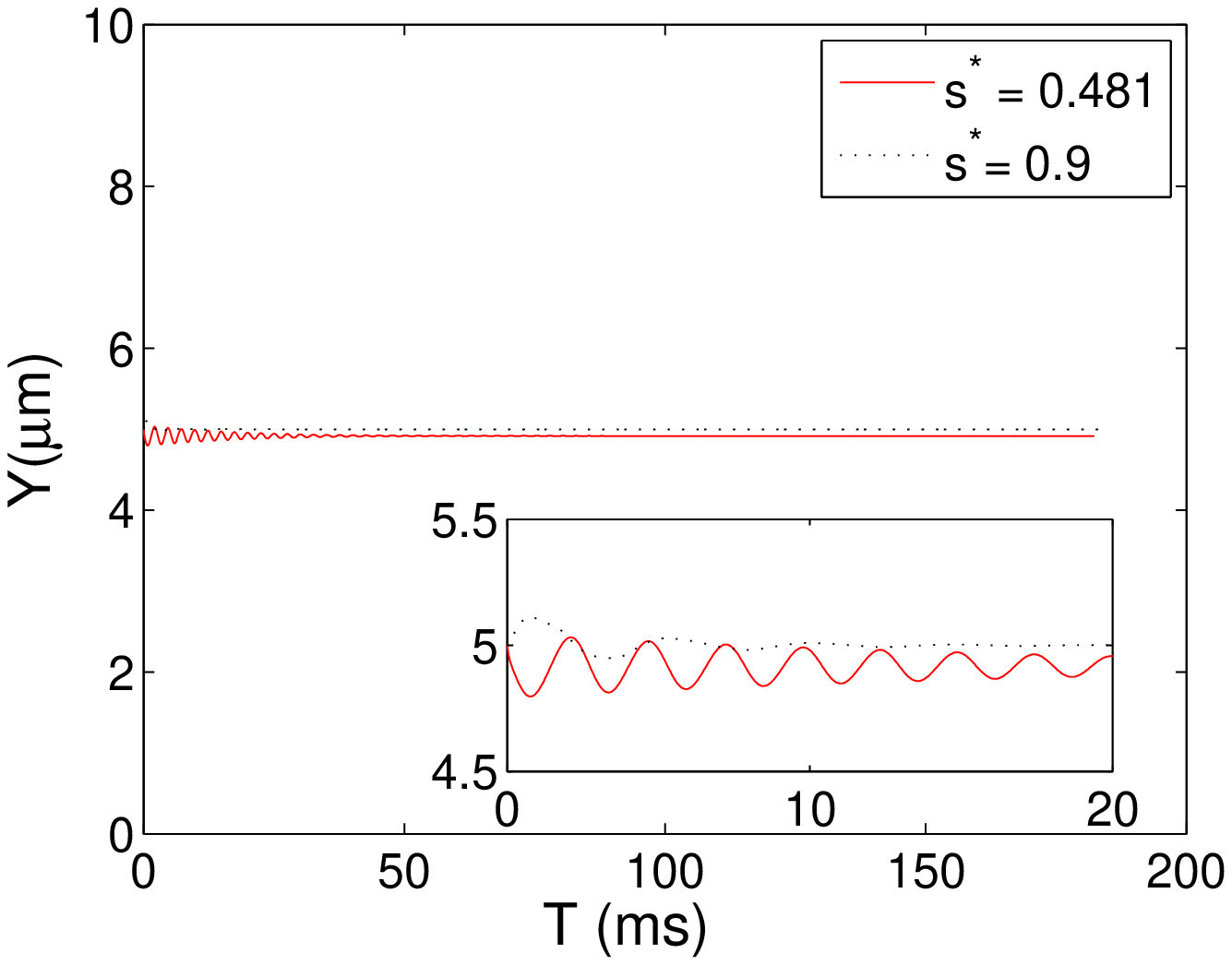}\hskip 5pt
\epsfxsize=2.5in \epsffile{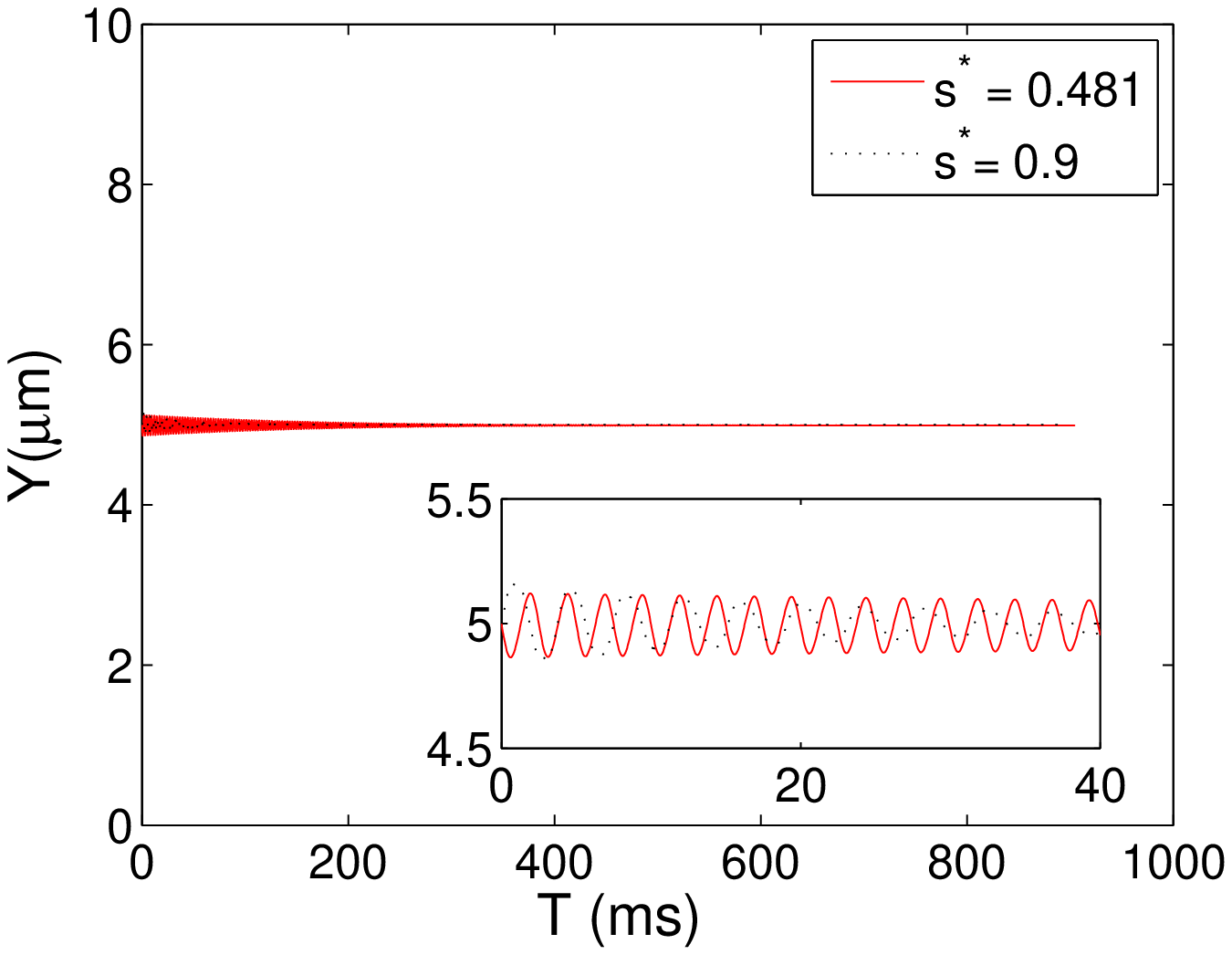}
\end{center}
\caption{(Color online). The history of the position of the cell mass center in Poiseuille flows for s*=0.481 and 0.9 with different bending constants: 10$k_b$ (left) and 100$k_b$ (right). $R_0/w =$ 0.56. 
The initial position is (5,5) at the channel centerline and the initial inclination angle is $\pi/4$.}
\label{101-umax075-history-45}
%\end{figure}
%\begin{figure}
\begin{center}
\leavevmode
\epsfxsize=2.5in \epsffile{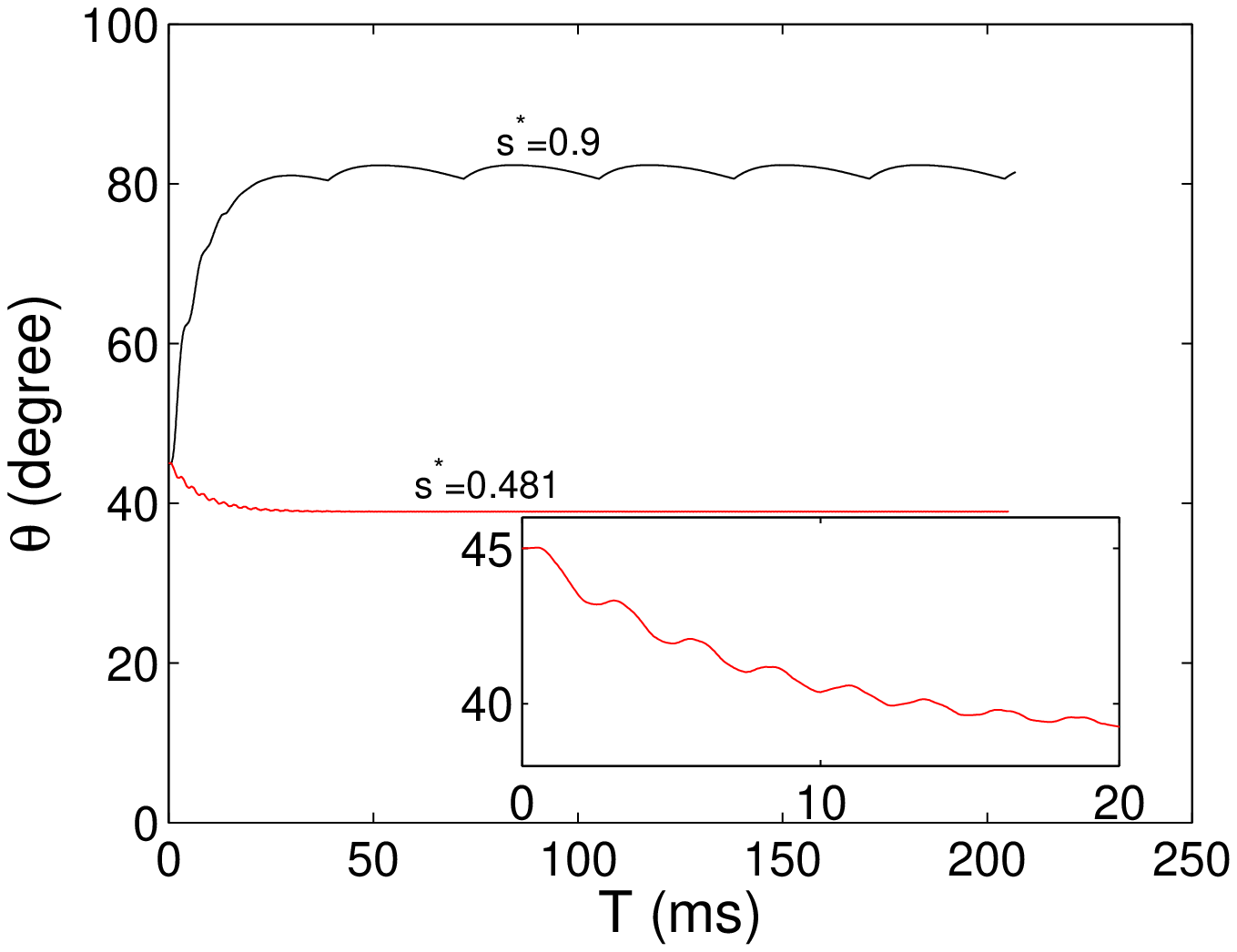}\hskip 5pt
\epsfxsize=2.5in \epsffile{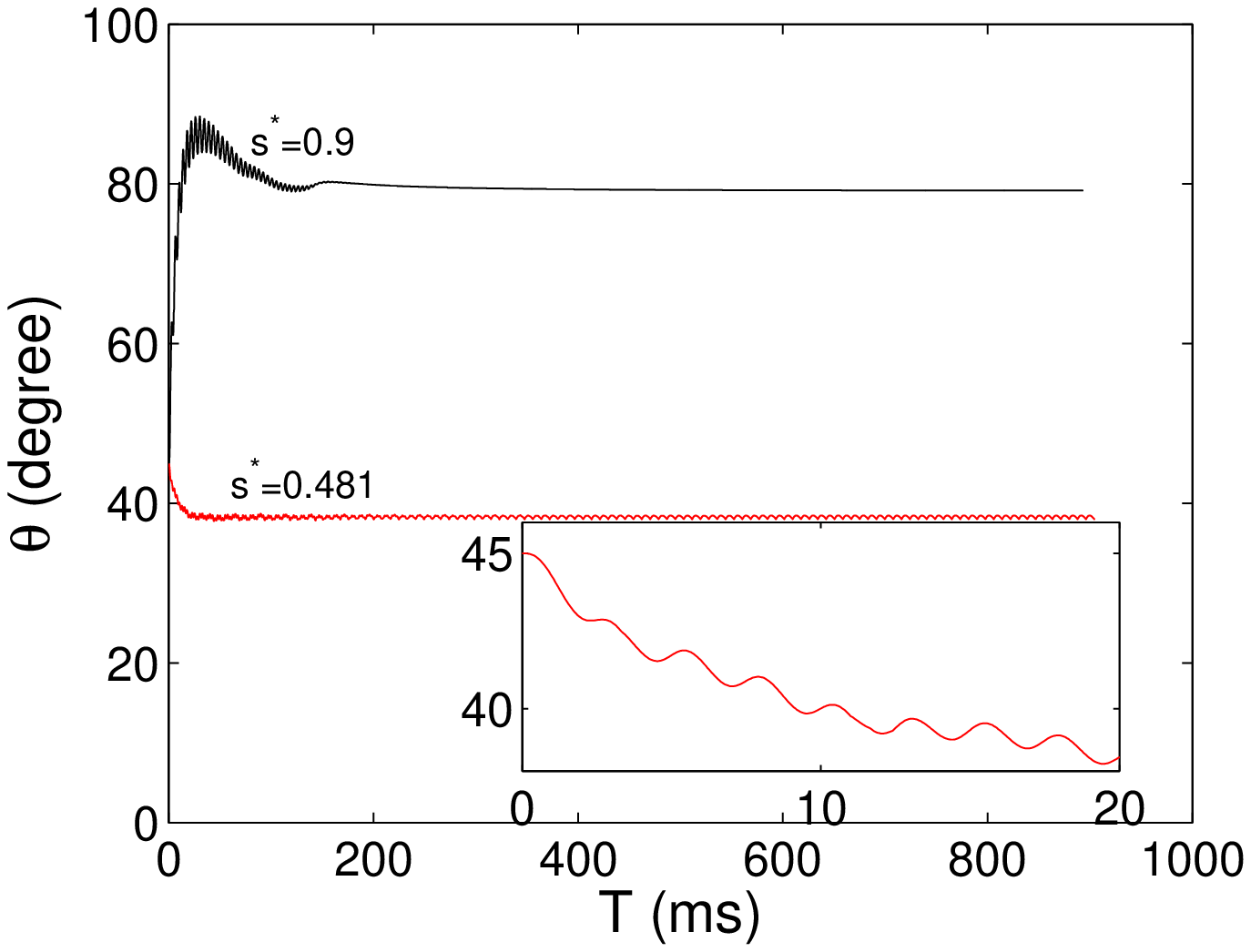}
\end{center}
\caption{(Color online). The history of the inclination angle $\theta$ between the long axis of cell and the horizontal line for the bending constants 10$kb$ (left) and 100$kb$ (right).  
The initial position is (5,5) at the channel centerline  and  the initial inclination angle is $\pi/4$.}
\label{101-umax075-angle-45}
\end{figure}

\begin{figure}
\begin{center}
\leavevmode
\epsfxsize=2.75in \epsffile{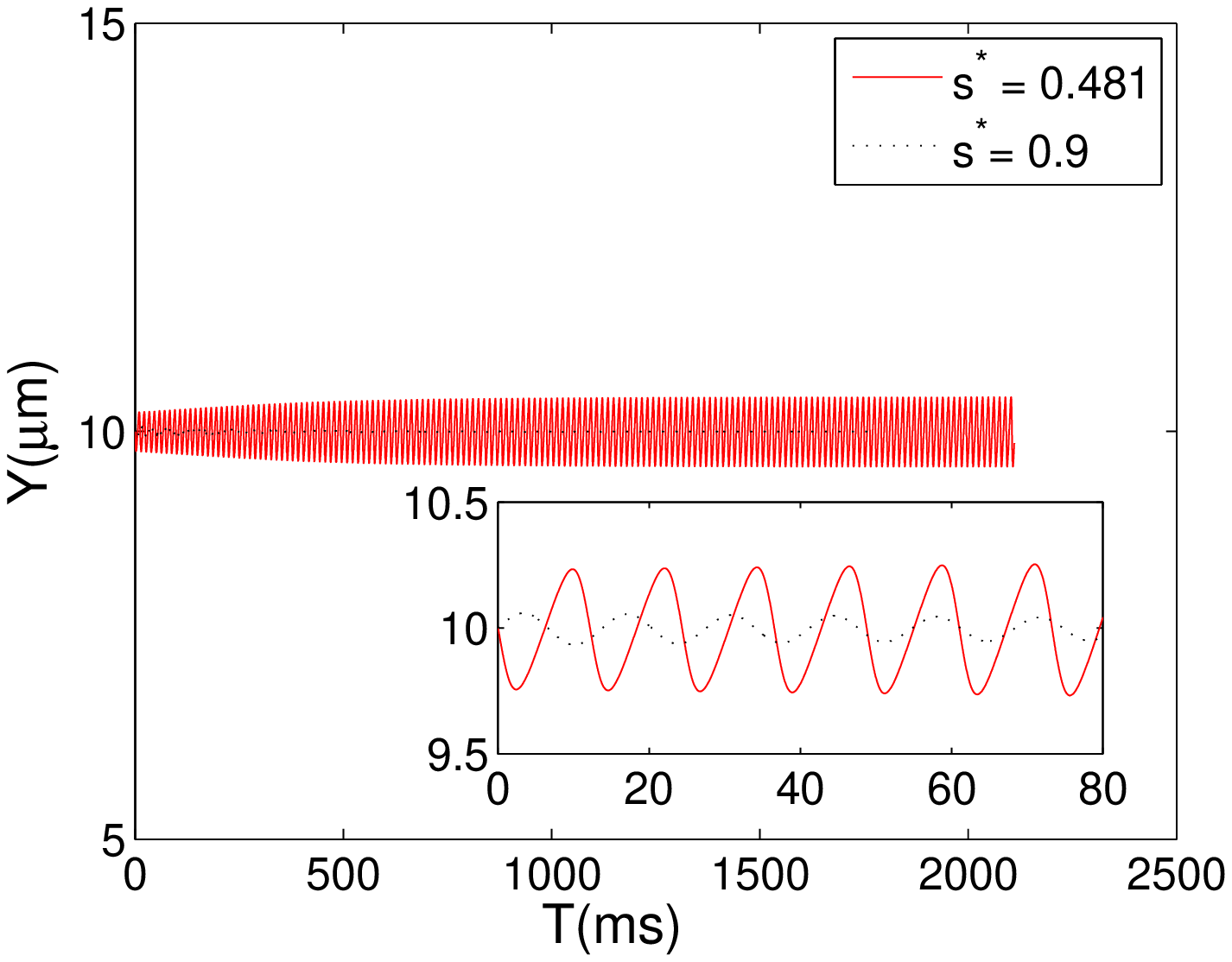}\hskip 5pt
\epsfxsize=2.75in \epsffile{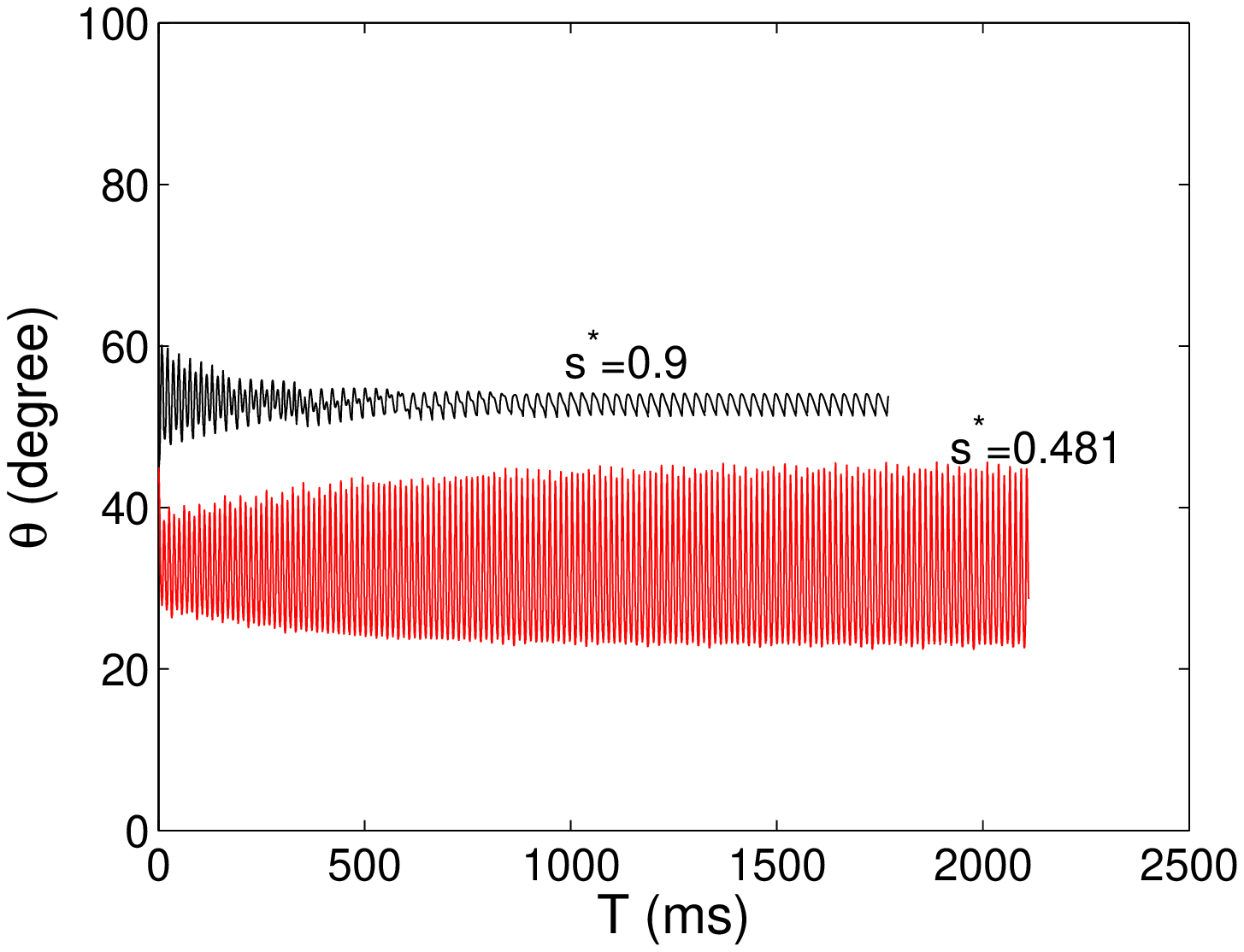}
\end{center}
\caption{(Color online). The history of the position of the cell mass center (left) and inclination angle (right) 
for s*=0.481 and 0.9 with the bending constant 100$k_b$ in a channel of height 20 $\mu$m. The initial position is 
(5,10)  at the channel centerline  and the initial inclination angle is $\pi/4$.}
\label{102-umax075-history-45}
%\end{figure}
%\begin{figure}
\begin{center}
\leavevmode
\epsfxsize=3in \epsffile{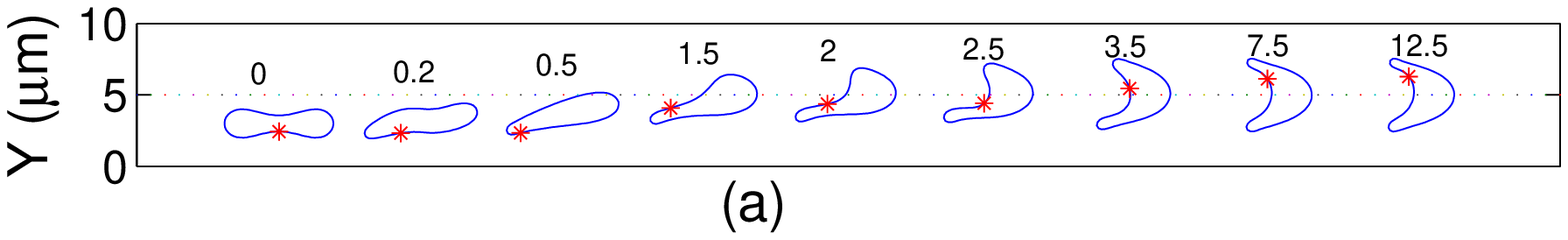}
\epsfxsize=3in \epsffile{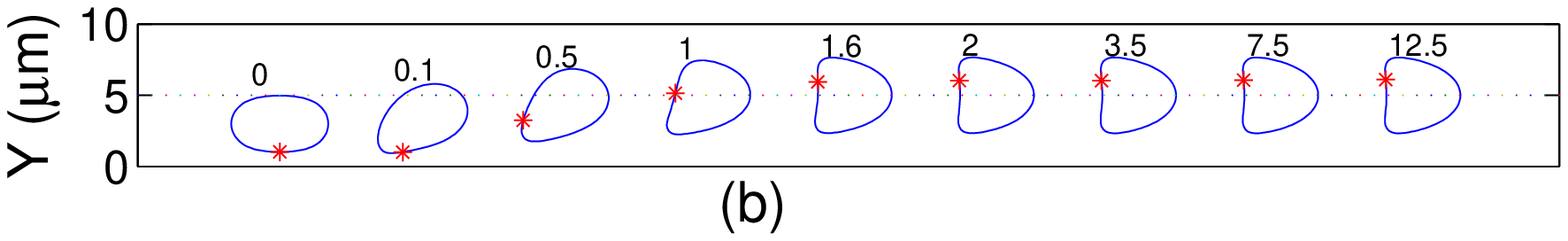}\\
\epsfxsize=3in \epsffile{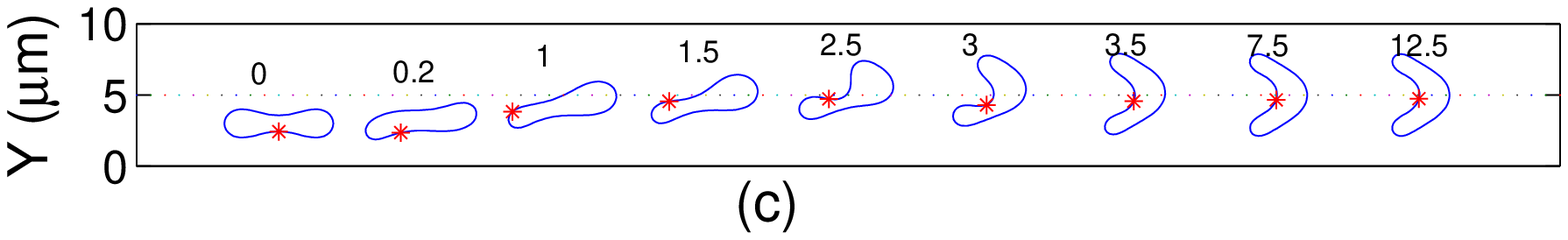}
\epsfxsize=3in \epsffile{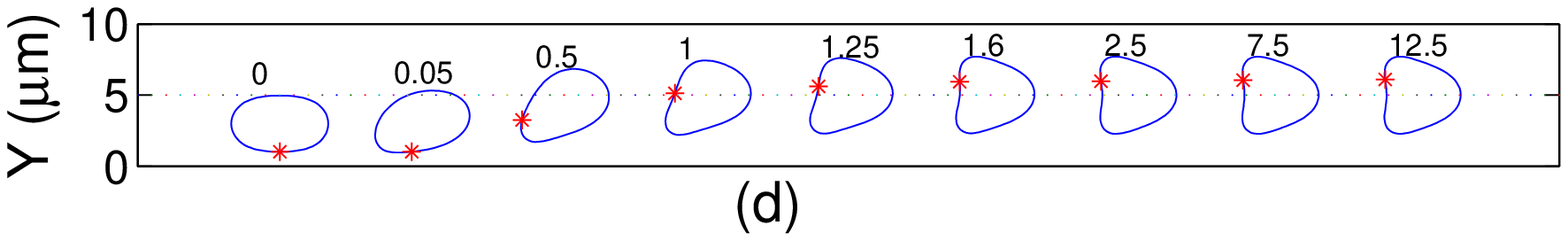}\\
\epsfxsize=3in \epsffile{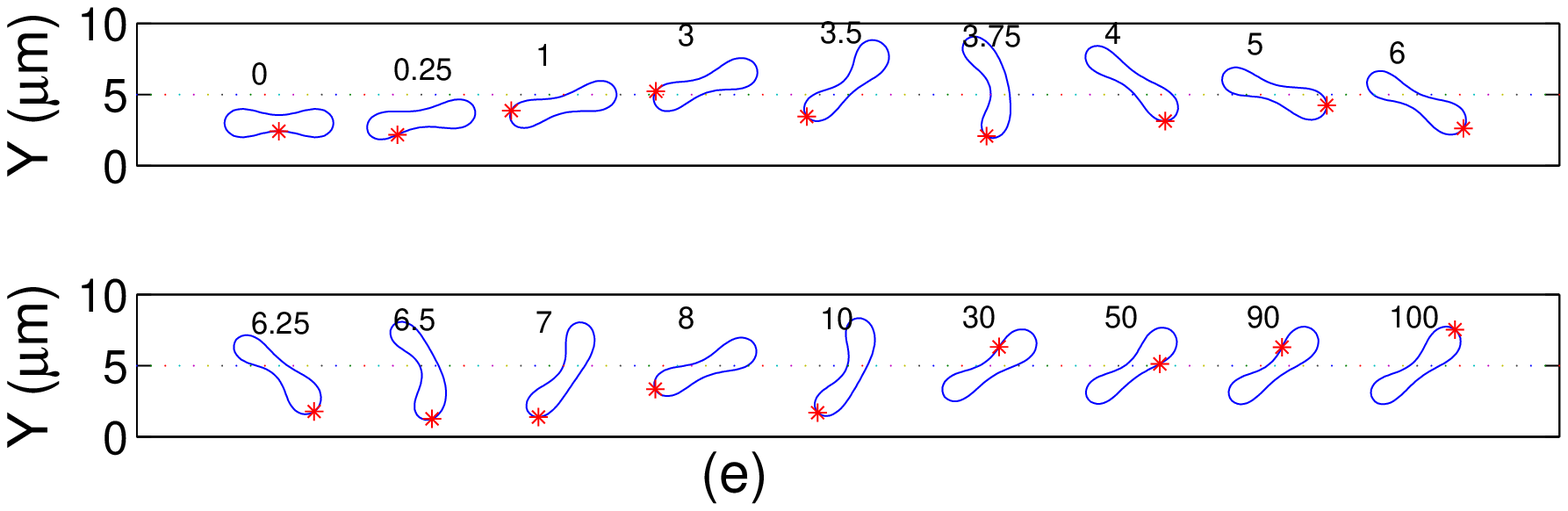}
\epsfxsize=3in \epsffile{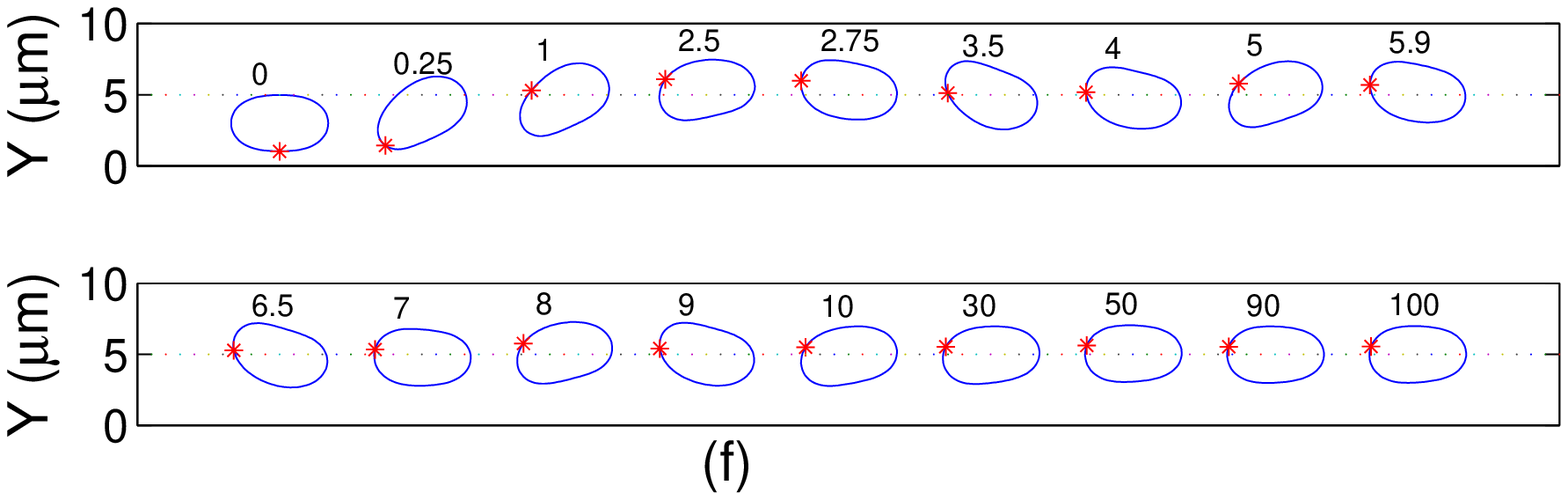}\\
\epsfxsize=3in \epsffile{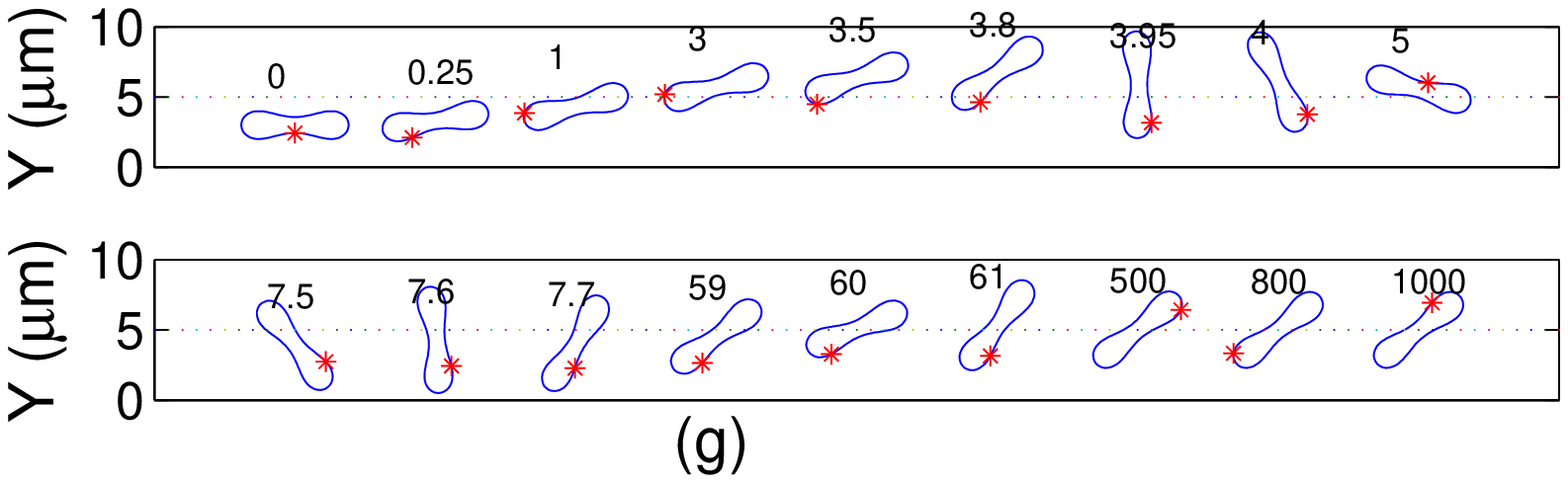}
\epsfxsize=3in \epsffile{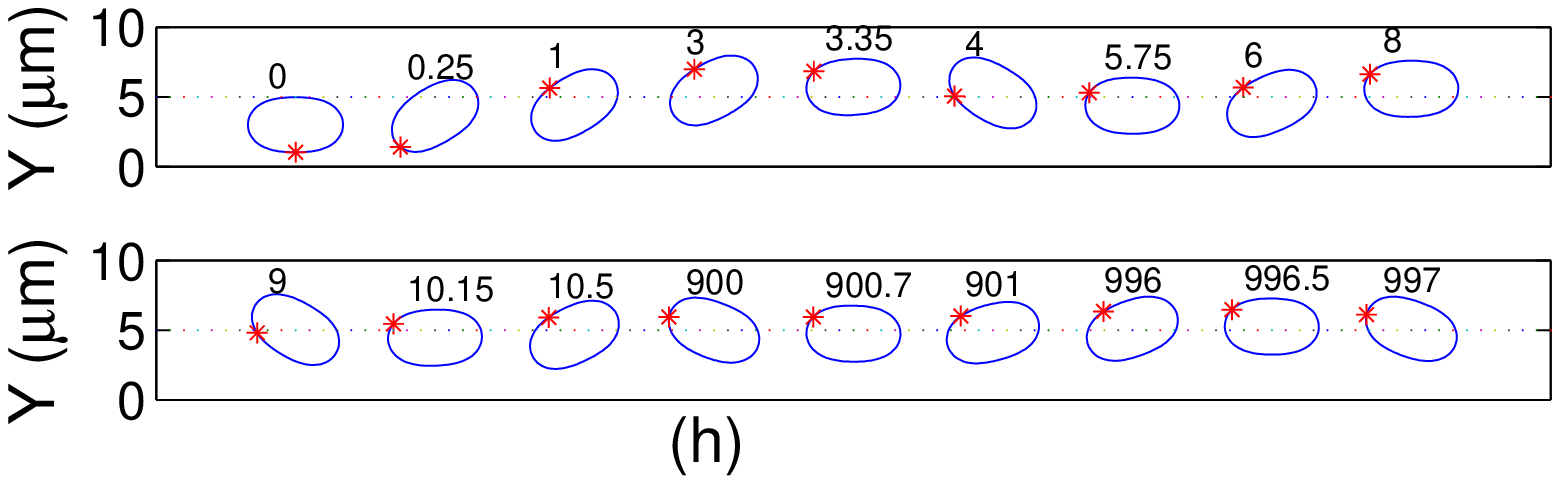}\\
\end{center}
\caption{(Color online). The snapshots of the cell migration in Poiseuille flows for $s^*$=0.481 and 0.9 with different bending
constants at different time ($\rm{ms}$): (a) $s^*$ = 0.481 and 0.1$k_b$, (b) $s^*$ = 0.9 and 0.1$k_b$,
(c) $s^*$ = 0.481 and 1$k_b$, (d) $s^*$ = 0.9 and 1$k_b$, (e) $s^*$ = 0.481 and 10$k_b$, (f) $s^*$ = 0.9 and 10$k_b$,
(g) $s^*$ = 0.481 and 100$k_b$, and (h) $s^*$ = 0.9 and 100$k_b$.
The red asterisk denotes the same node point on the cell membrane. 
The initial position is (5,3) and the initial inclination angle is 0.}
\label{101-umax075-snapshot}
\end{figure}

Two motions of oscillation and vacillating-breathing (swing) of an RBC are observed in a 
narrow (100 $\times$ 10 $\rm{\mu m}^2$) channel considered here.
The values of parameters for modeling cells are same with \cite{Shi2012a,Shi2012b} 
as follows: The bending constant is $k_{b}=5\times 10^{-10}$ $\rm{N m}$, the spring constant is
$k_{l}=5\times10^{-8}$ $\rm{N m}$, and the penalty coefficient is $k_s= 10^{-5}$ $\rm{N m}$.
The swelling ratios of the cells in the simulations are $s^*$ = 0.481 and 0.9.
The cells are suspended in blood plasma which has a density $\rho=1.00$ $\rm{g/cm^{3}}$ and a dynamical
viscosity  $\mu$ = 0.012 $\rm{g/(cm s)}$. The viscosity ratio which describes the viscosity contrast
of the inner and outer fluid of the RBC membrane is fixed at 1.0. The computational domain is a two
dimensional horizontal channel.  In addition, periodic conditions are imposed at the left and right 
boundary of the domain. The Reynolds number is defined by $Re = {\rho}UH/\mu$, where $U$ is the 
average velocity in the channel, and $H$ is the height of the channel.
To obtain a Poiseuille flow, a constant pressure gradient is prescribed
as a body force  so that the Reynolds number of the Poiseuille flow without cell is about 0.4167. 
We consider the motion of a single RBC with four different bending constants which are 
0.1$k_b$, 1$k_b$, 10$k_b$, and 100$k_b$ in a Poiseuille flow with the fluid domain 100 $\times$ 10 $\rm{\mu m}^2$ 
(the degree of confinement $R_0/w$ = 0.56).  The initial velocity is zero everywhere. The grid resolution for 
the computational domain is 64 grid points per 10 $\rm{\mu m}$. 

%The capillary numbers $C_a$ are 268.53, 26.853, 2.685 and 0.268 corresponding to the bending
%constants 0.1$k_b$, 1$k_b$, 10$k_b$, and 100$k_b$, respectively. The capillary number is defined by
%$C_a = \mu {G_r} {R_0}^3 / {B}$, where $\mu$, ${G_r}$, ${R_0}$ and ${B}$ represent the plasma viscosity, 
%the shear rate of fluid flow  based on the gradient of the velocity at the wall, the effective radius of the cell, 
%and the bending coefficient, respectively.

First we have studied the cases in which the initial position of the cell mass center is located at the centerline  and the initial inclination angle is $\theta=\pi/4$.
Different motions led by the different bending constants are observed.
When the bending constants are 0.1$k_b$ and 1$k_b$, the cells of both swelling ratios stay at
the center region of the channel and the parachute shapes have been obtained for both cells as shown in 
Figures \ref{101-umax075-snapshot-45} $(a)$-$(d)$.
For the bending constant 10$k_b$, the cell of swelling ratio $s^*$ = 0.481 exhibits a damped 
oscillation with deformation (called the vacillating-breathing behavior \cite{Misbah2006}) 
until it attains the equilibrium state aligning itself at an angle
with the direction of the flow as shown in Figures \ref{101-umax075-snapshot-45} $(e)$ and \ref{101-umax075-angle-45}. 
The cell has a slipper shape whose mass center is slightly away from the center line
as studied in \cite{Shi2012b}.
But the one of swelling ratio $s^*$ = 0.9 only has oscillating motion for shorter period of time
(see Figure \ref{101-umax075-snapshot-45} $(f)$) and the cell gradually deforms into a parachute shape.
When the bending constant is 100$k_b$, both cells first exhibit damped 
oscillation with the shapes of long body and, at the end, align themselves with a fixed inclination angle with respect
to the flow direction as in Figures \ref{101-umax075-snapshot-45} $(g)$, $(h)$ and \ref{101-umax075-angle-45}.
The histories of the height of the cell mass center in Figure \ref{101-umax075-history-45}
do show the correlation with the oscillating motion.
The initial inclination angle $\theta=\pi/4$ helps the fluid flow to create the oscillation of the cell mass center 
and then the flow field, which is about a full quadratic profile, interacting with the 
long body creates the oscillation similar to the one in \cite{Sugihara-Seki1993}.
But the cell mass center gradually approaches to an equilibrium height due to the interaction 
between the deformability of the cell and the Poiseuille flow and the wall effect 
as in Figure \ref{101-umax075-history-45} and at the 
same time the oscillating motion is damping out accordingly as shown
by the histories of the inclination angle in Figure \ref{101-umax075-angle-45}.
Once the oscillation of the cell mass center damps out, the inclination angle then is fixed as 
in Figures \ref{101-umax075-history-45}  and \ref{101-umax075-angle-45}.

When increasing the channel height to 20 $\rm{\mu m}$  and 
keeping the other parameters same, we have focused now on the cases of the bending constant  
100$k_b$. In Figure \ref{102-umax075-history-45}, the histories show that
the oscillation of inclination angle become periodic. These cells with the bending constant  
100$k_b$ stay at the channel central region since they are not neutrally buoyant rigid particle and
and the deformability plus the Poiseuille flow profile keeps them staying in the channel central region.
But the wall effect is weaker in a twice wider channel so that the up and down motion of the mass center
does not damps out fast as in the channel of height 10 $\rm{\mu m}$.

\begin{figure}
\begin{center}
\leavevmode
\epsfxsize=2.5in \epsffile{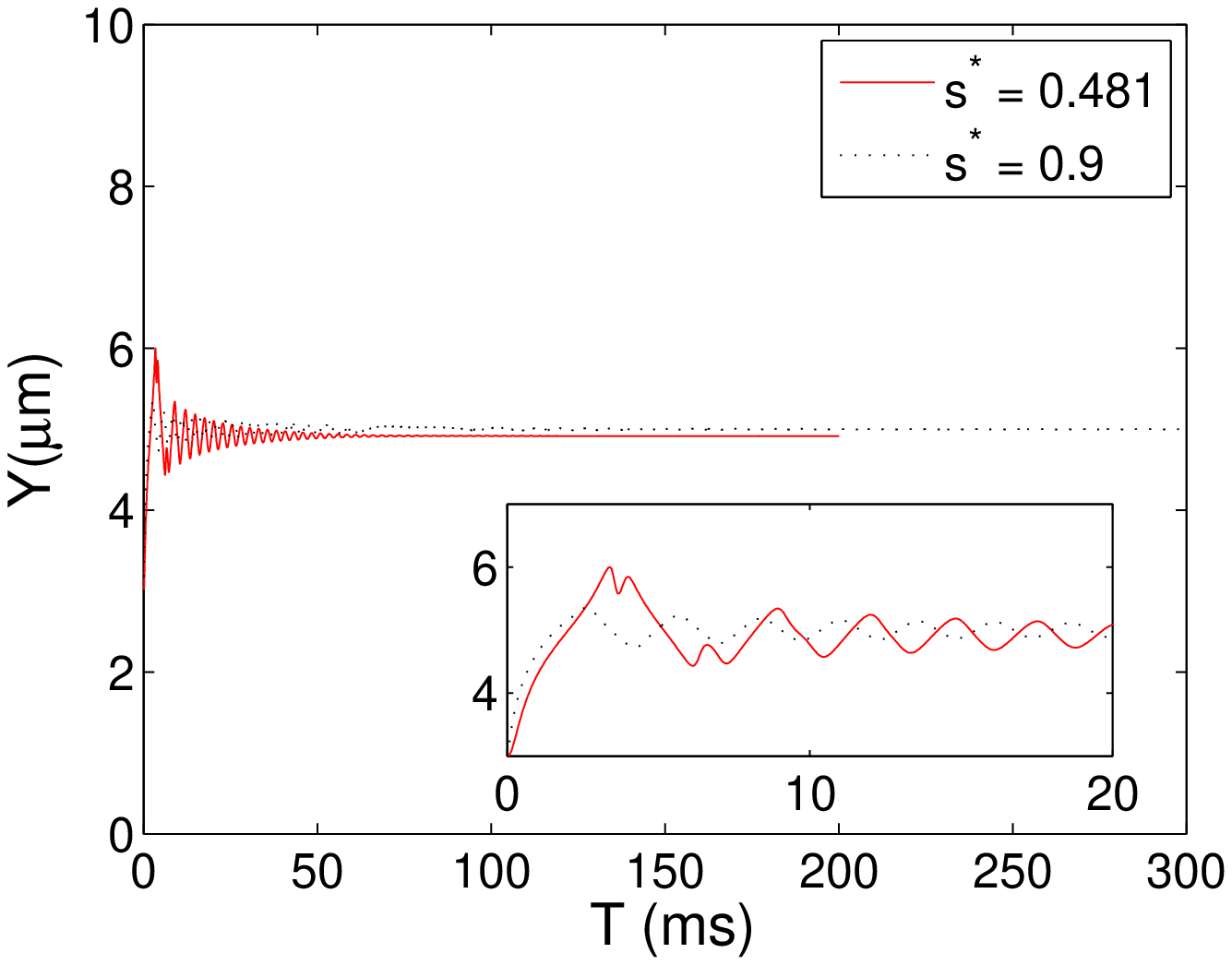}\hskip 5pt
\epsfxsize=2.5in \epsffile{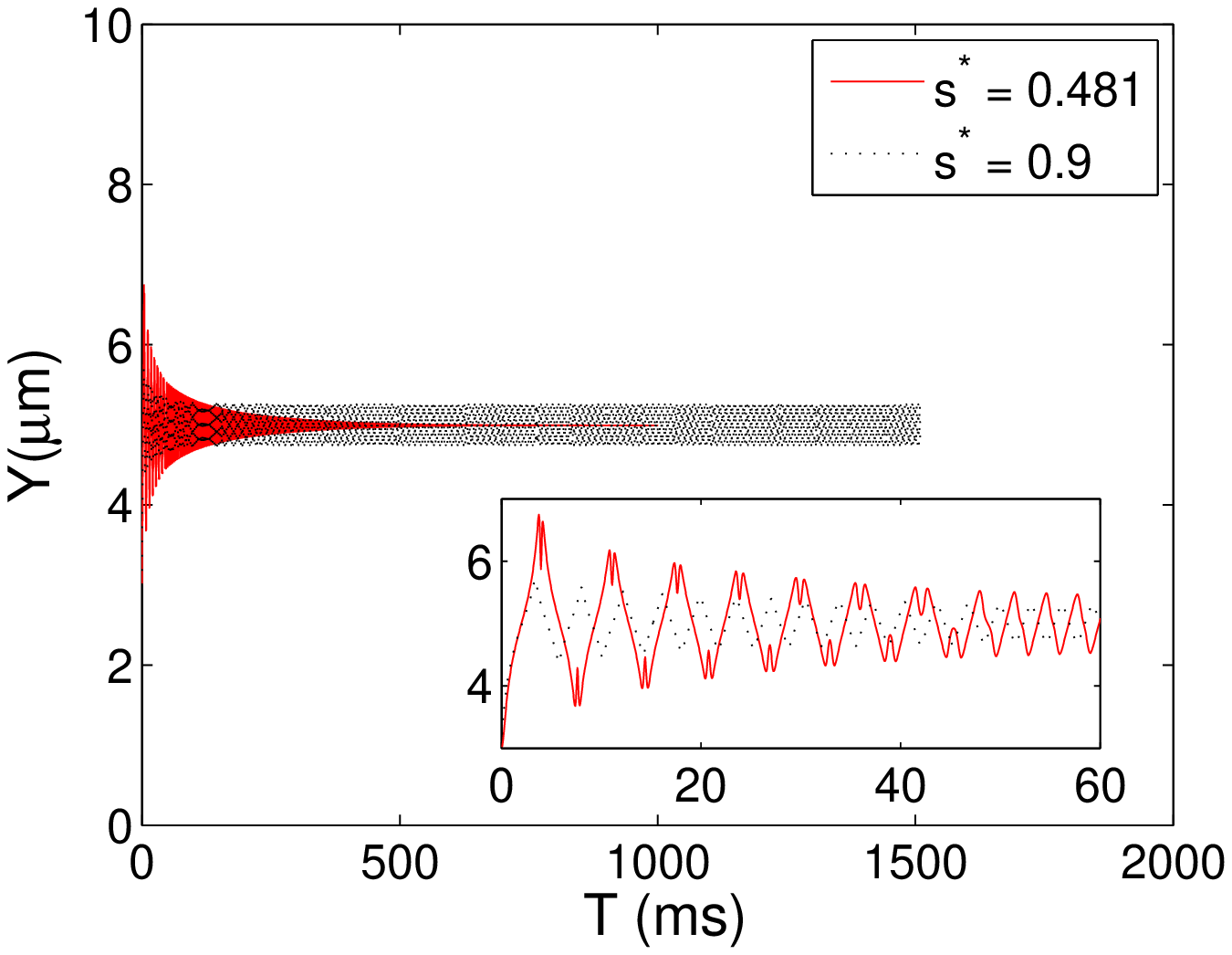}
\end{center}
\caption{(Color online). The history of the position of the cell mass center in Poiseuille flows for s*=0.481 and 0.9 with different bending constants: 10$k_b$ (bottom left), and 100$k_b$ (bottom right). 
The initial position is (5,3) and the initial  inclination angle is 0.}
\label{101-umax075-history}
%\end{figure}
%\begin{figure}
\begin{center}
\leavevmode
\epsfxsize=2.5in \epsffile{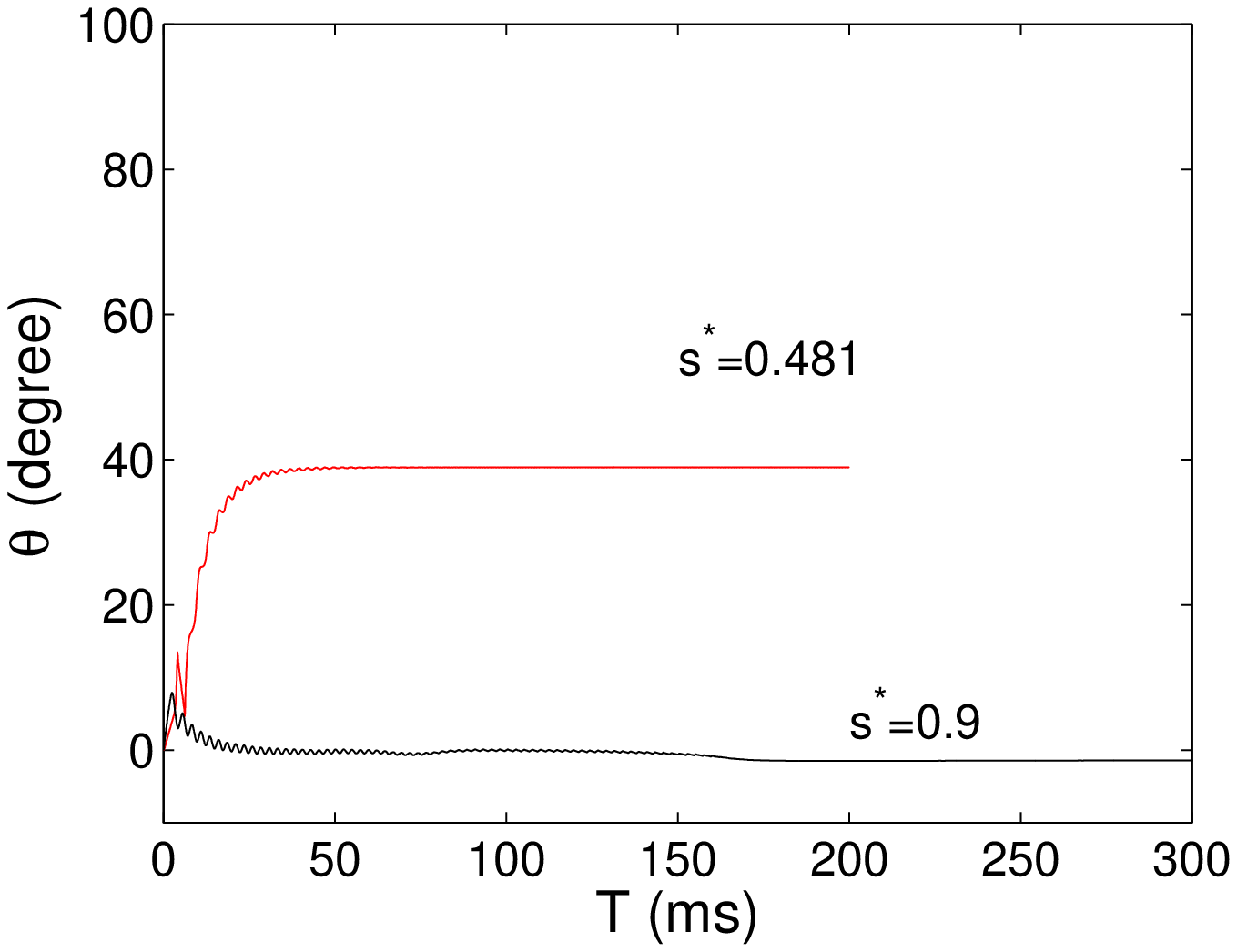}\hskip 5pt
\epsfxsize=2.5in \epsffile{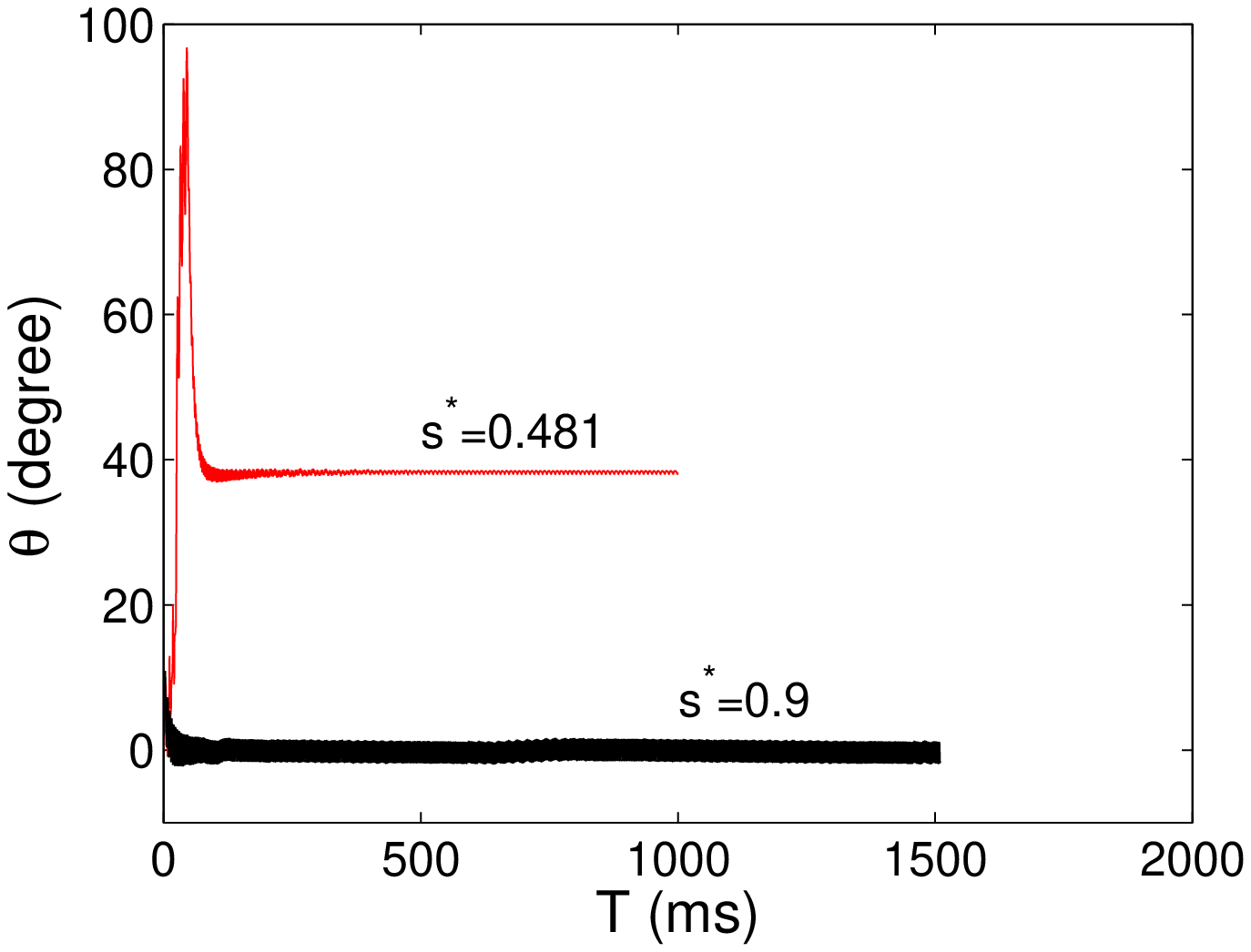}
\end{center}
\caption{(Color online). The history of the angle between the long axis of cell and the horizontal line for the bending constants 10$kb$ (left) and 100$kb$ (right).
The initial position is (5,3) and the initial  inclination angle is 0.}
\label{101-umax075-angle}
\end{figure}

When placing the initial position of the cell mass center off the centerline in a channel
of height of 10 $\rm{\mu m}$ , we have obtained almost similar behaviors
for both cells in Poiseuille flow as shown in Figures \ref{101-umax075-snapshot}, \ref{101-umax075-history}
and \ref{101-umax075-angle}. The vacillating-breathing behavior for the case of the bending constant 10$k_b$
and the oscillating motion for the case of the bending constant 100$k_b$  are  stronger for the 
cell of $s^*=0.481$ since the lateral migration toward the channel central region enhances the oscillation. 
But another interesting result is the motion of the cell of $s^*=0.9$ for
the bending constants 10$k_b$ and 100$k_b$ as shown in Figures \ref{101-umax075-snapshot}  $(f)$ and $(h)$.
A similar motion called snaking motion in Poiseuille flow has been studied in the Stokes regime in \cite{Koui2011}.  

Concerning the oscillating motion and vacillating-breathing motion in Poiseuille flow, the
bending constant needs to be large enough with respect to the maximum velocity of the fluid flow $u_\textrm{max}$
so that the cell can not be deformed into a symmetric parachute in the channel central region, 
but maintain a long body shape to interact with Poiseuille flow.

\subsection{Oscillating motion of a neutrally buoyant particle in a narrower channel}\label{sec.3.2}

\begin{figure}
 \begin{center}
    \leavevmode
    \epsfxsize=2.9in \epsffile{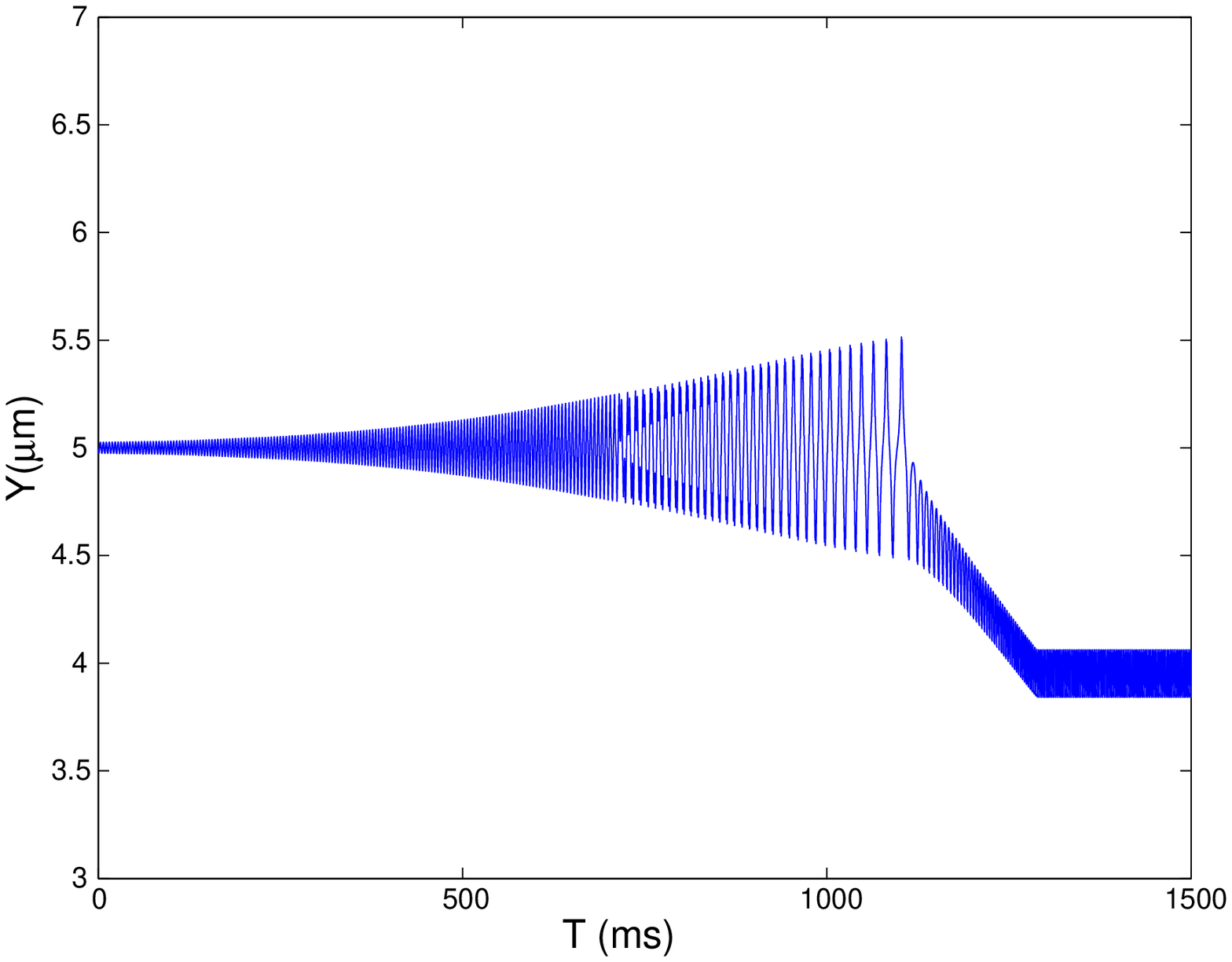}
    \epsfxsize=2.9in \epsffile{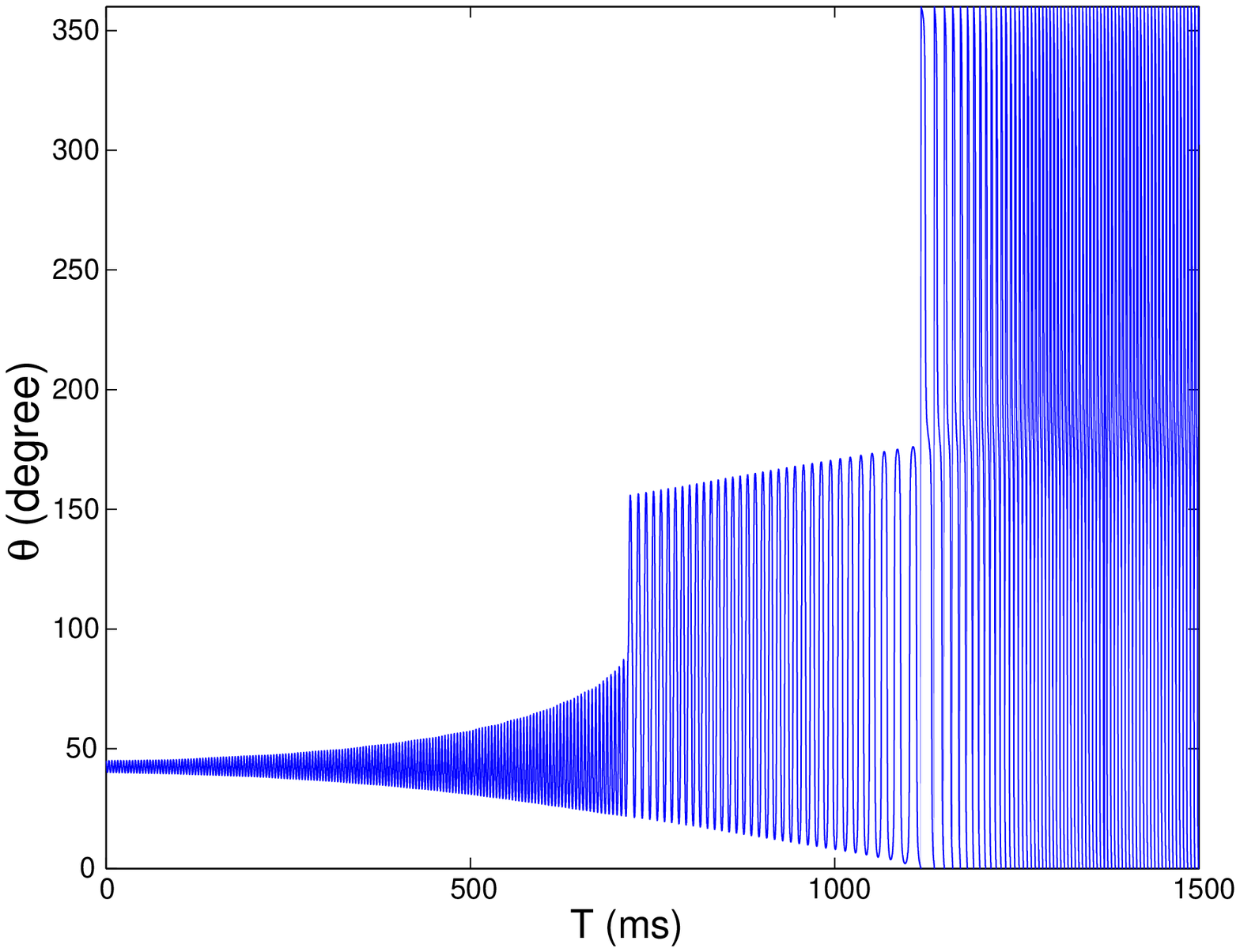}\\
    \epsfxsize=2.9in \epsffile{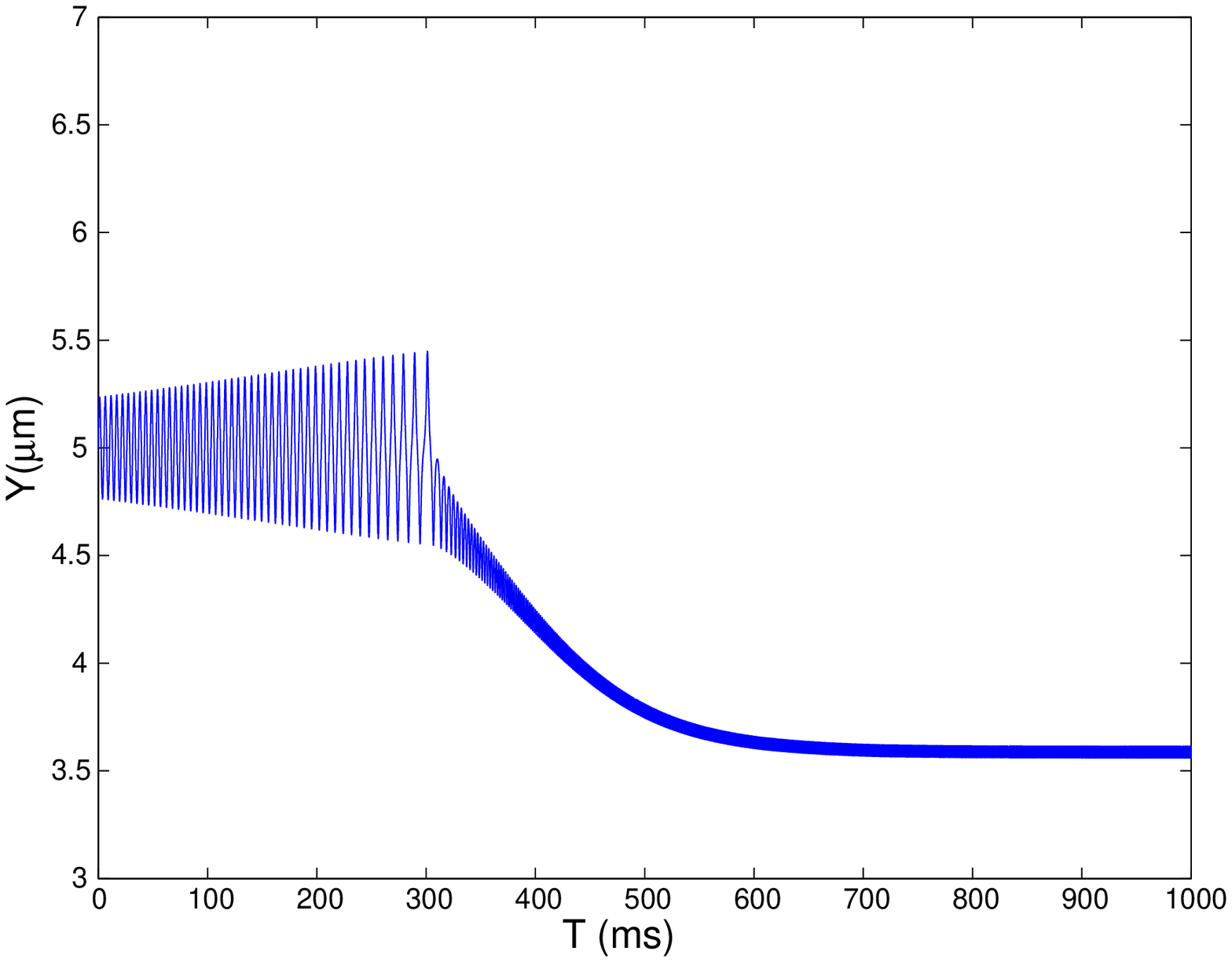}
    \epsfxsize=2.9in \epsffile{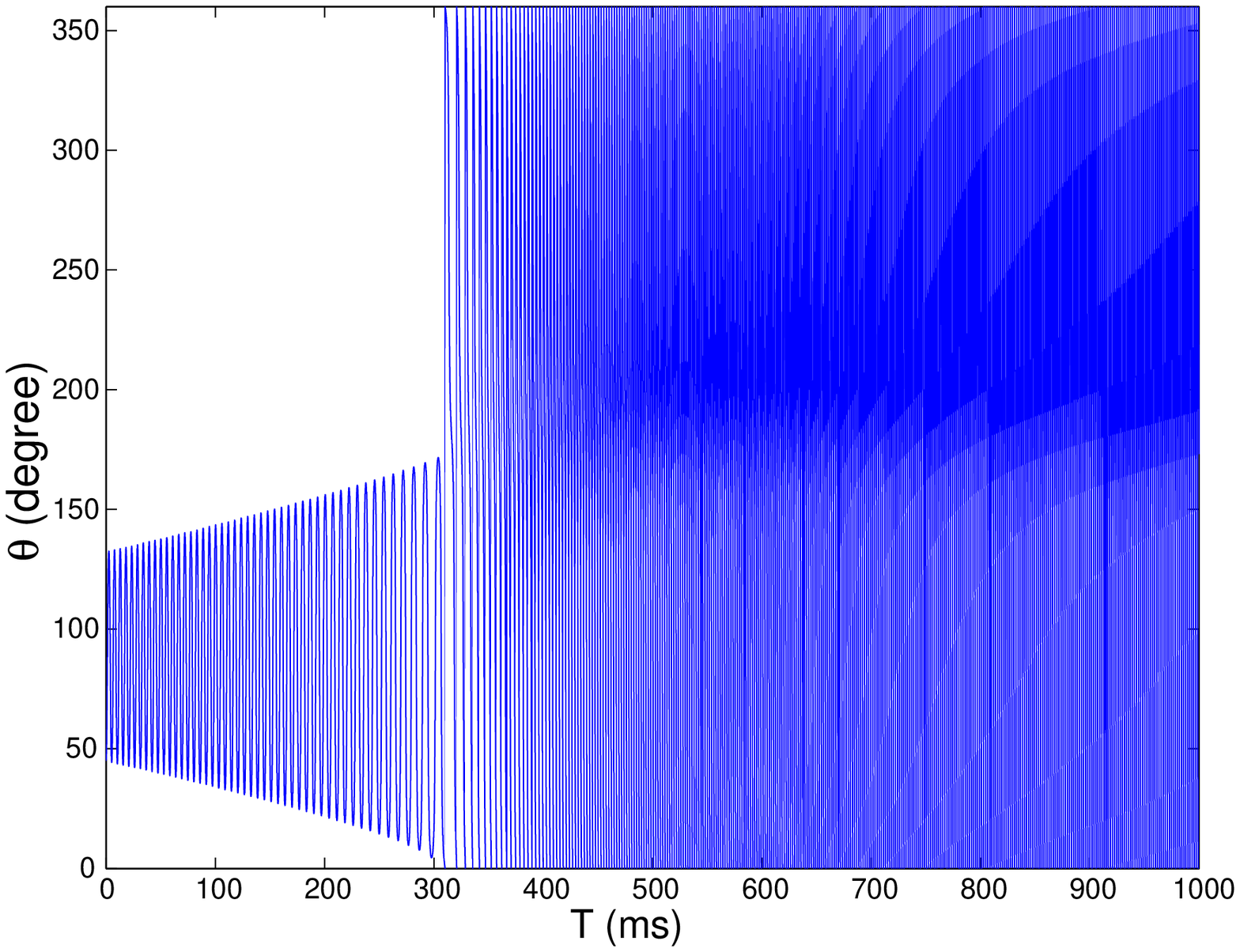}\\
 \end{center}
 \caption{(Color online). History of particle mass center and inclination angle of the biconcave
particle with $s^*=0.481$(top two) and the elliptic particle with $s^*=0.9$ (bottom two).}\label{ptk.ctr.y1}
\end{figure}
\begin{figure}
 \begin{center}
    \leavevmode
     \epsfxsize=5in \epsffile{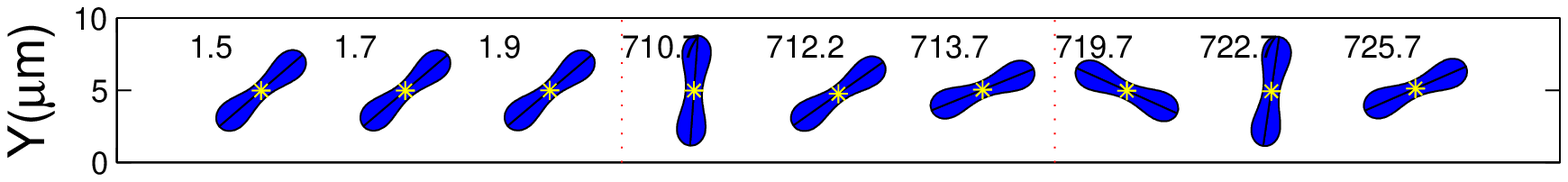}\\
    \epsfxsize=5in \epsffile{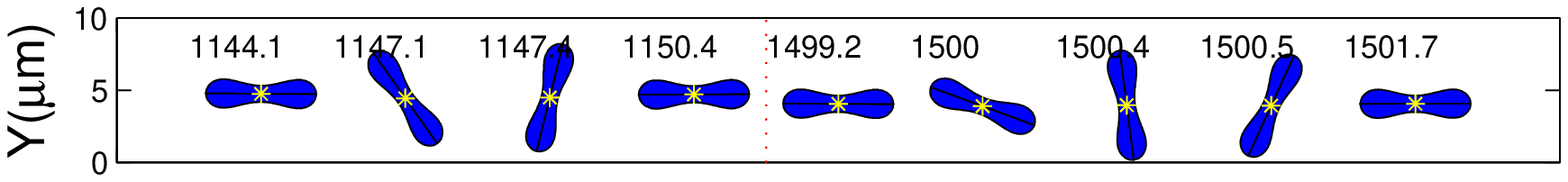}\\
    \epsfxsize=5in \epsffile{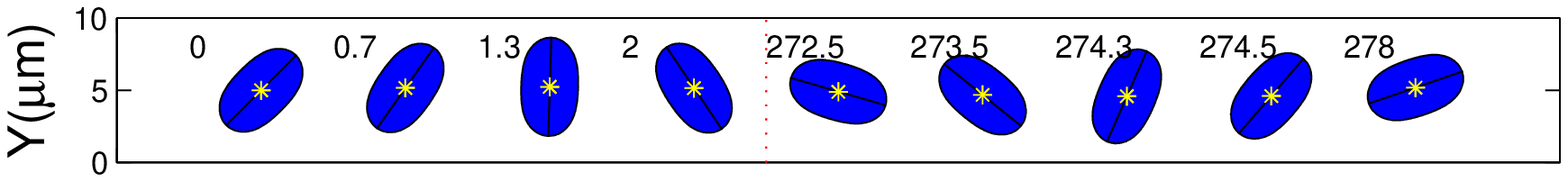}\\
    \epsfxsize=5in \epsffile{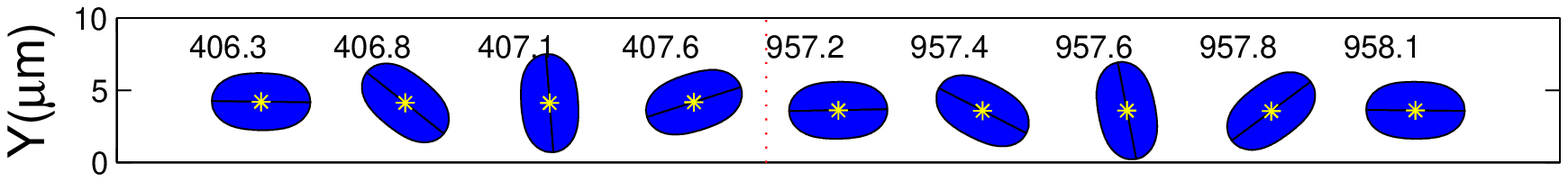}\\
 \end{center}
 \caption{(Color online). Snapshots of particle motion in a narrow channel with the initial position of the mass center  at
the channel centerline. (i). Biconcave shape: $0^o < \theta < 90^o$ for $1.5 \le t \le 1.9$ and 
$710.7 \le t \le 713.7$, $0^o < \theta < 180^o$ for $719.7 \le t \le 725.7$  and  $1144.1 \le t \le 1150.4$, 
and tumbling for  $1499.2 \le t \le 1501.7$. (ii) Elliptical shape: $0^o < \theta < 180^o$ for $0 \le t \le 2$
and $272.5 \le t \le 278$ and tumbling  $406.3 \le t \le 407.6$ and $957.2 \le t \le 958.1$.
The number above each particle indicates the time (time unit is ms).}\label{snapshot.ctr.y1}
\end{figure}

\begin{figure}
 \begin{center}
    \leavevmode
    \epsfxsize=2.9in \epsffile{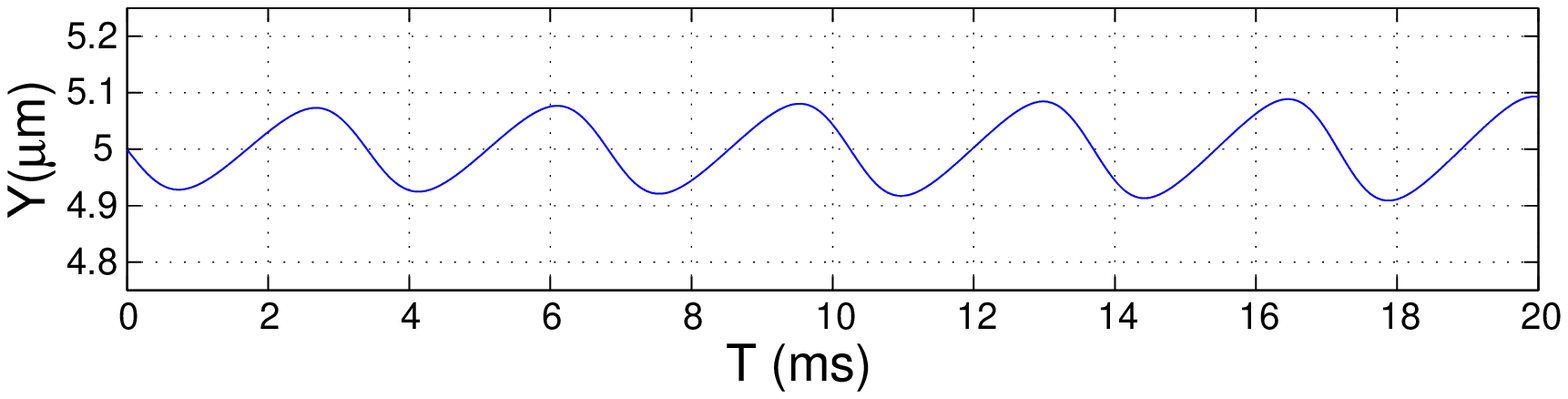}\hskip 5pt \epsfxsize=2.9in \epsffile{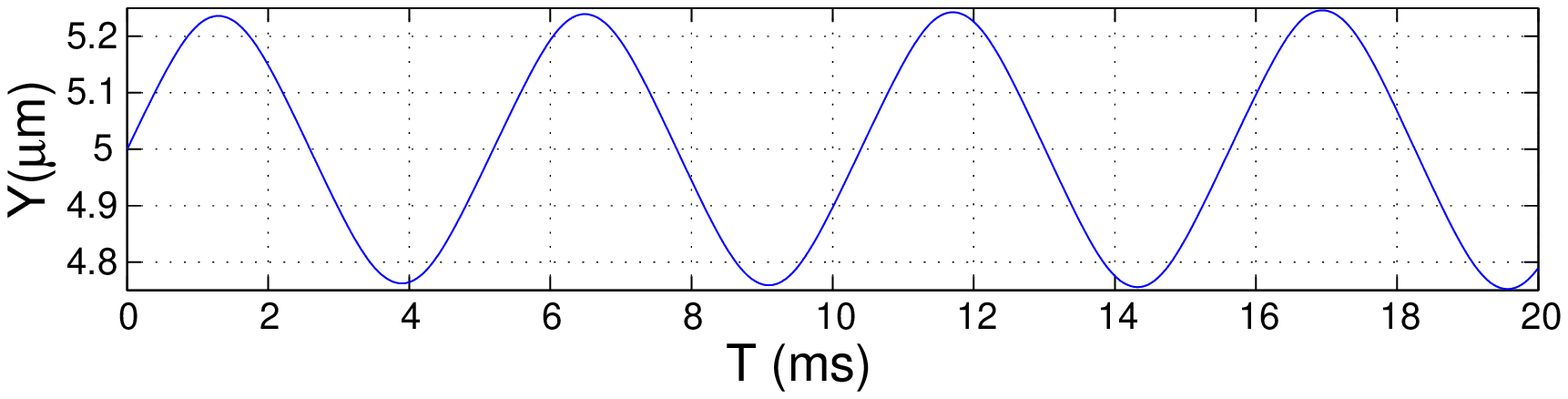}\\
     \epsfxsize=2.9in \epsffile{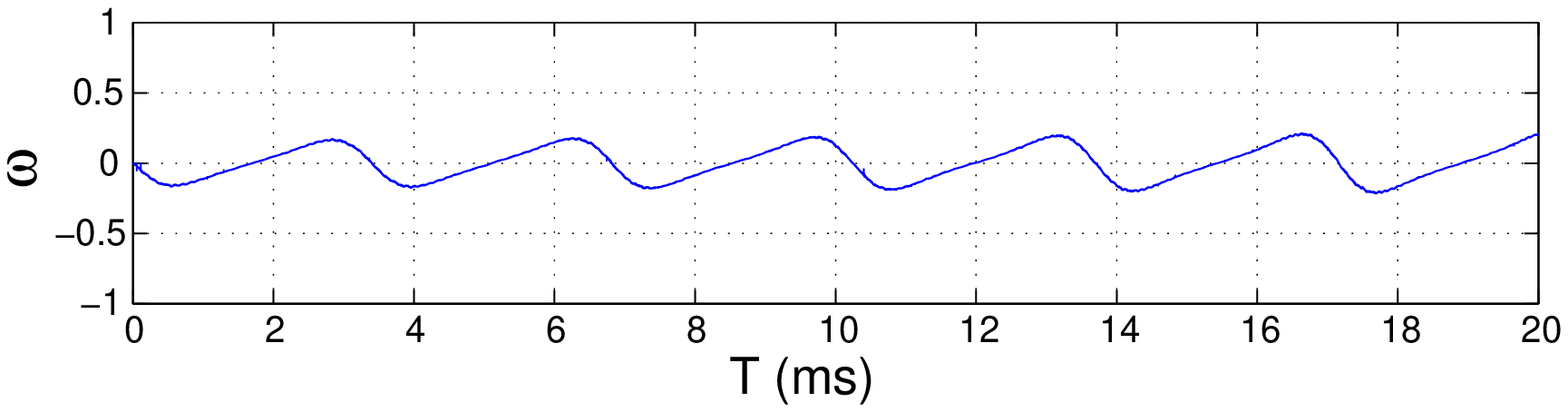}\hskip 5pt\epsfxsize=2.9in \epsffile{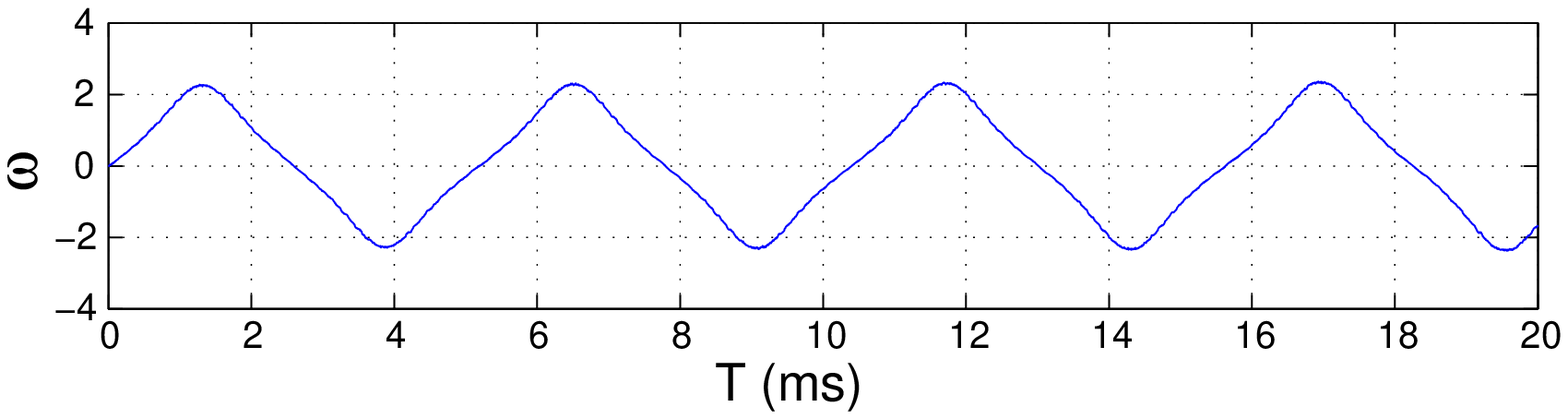}
 \end{center}
 \caption{(Color online). The histories of  particle mass position (top) and the angular velocity (bottom) with $s^*=0.481$ (left two)
 and  $s^*=0.9$ (right two) for $0 \le t \le 20$. }\label{ang.vs.ycr}
\end{figure}

\begin{figure}
 \begin{center}
    \leavevmode
    \epsfxsize=2.9in \epsffile{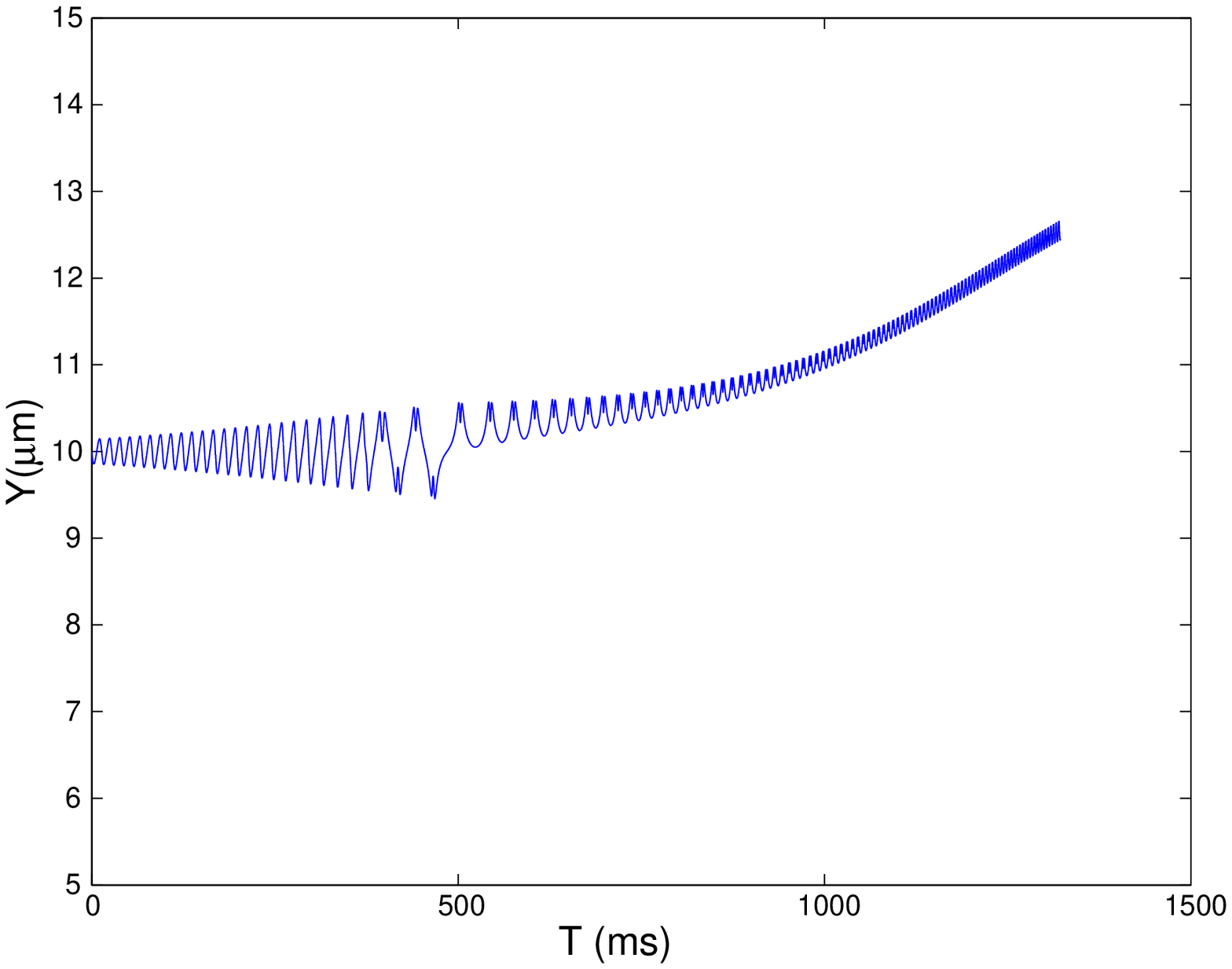}
    \epsfxsize=2.9in \epsffile{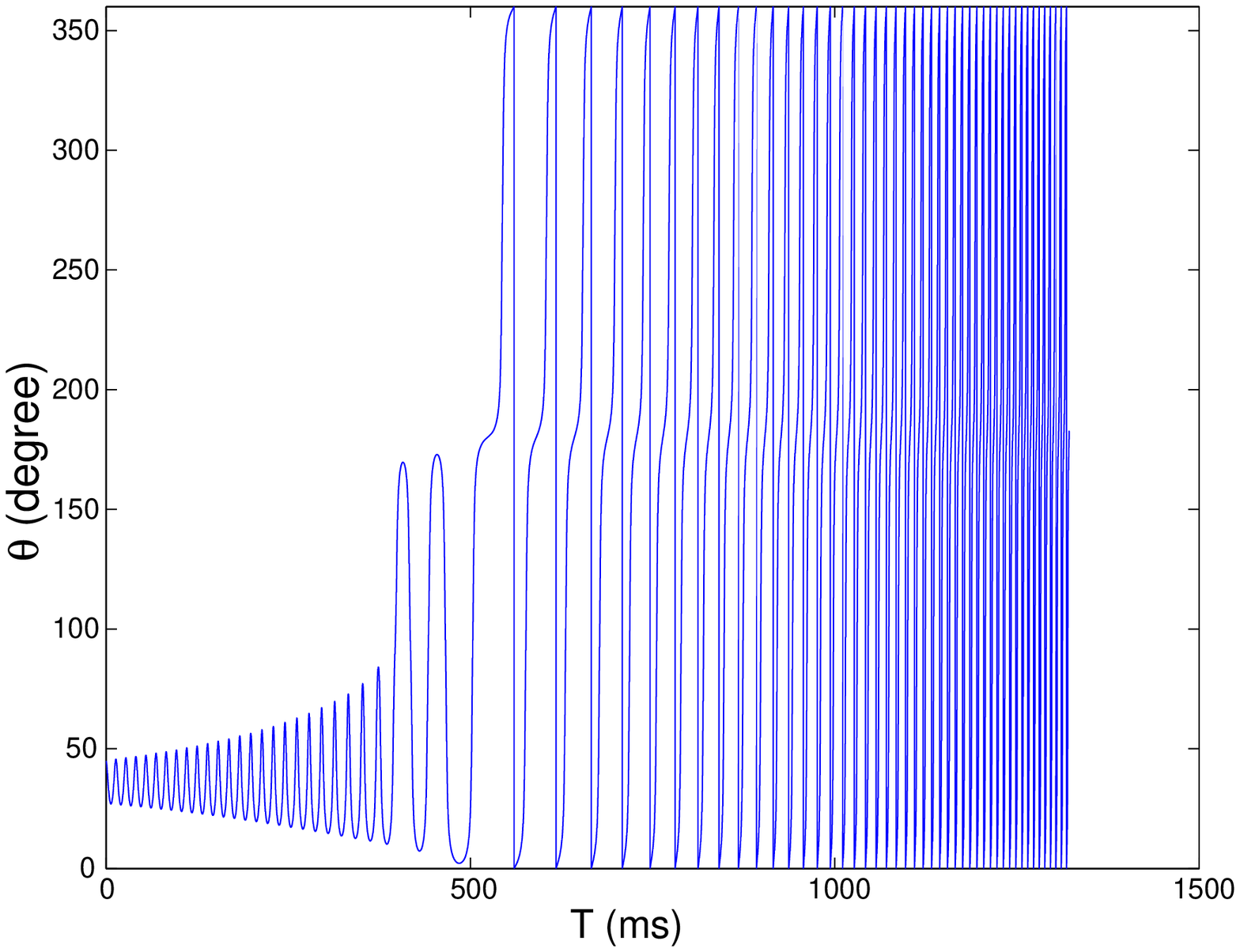}\\
    \epsfxsize=2.9in \epsffile{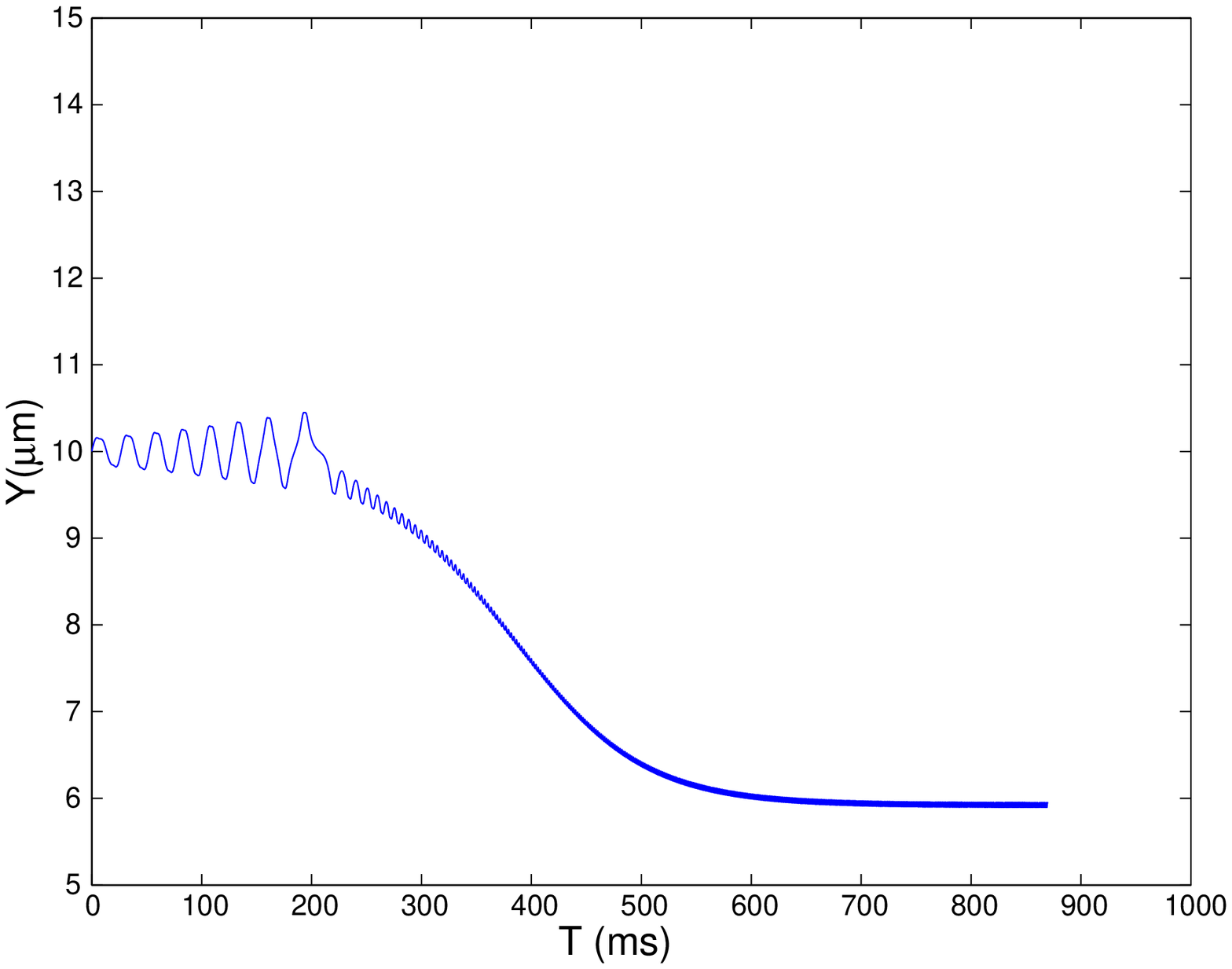}
    \epsfxsize=2.9in \epsffile{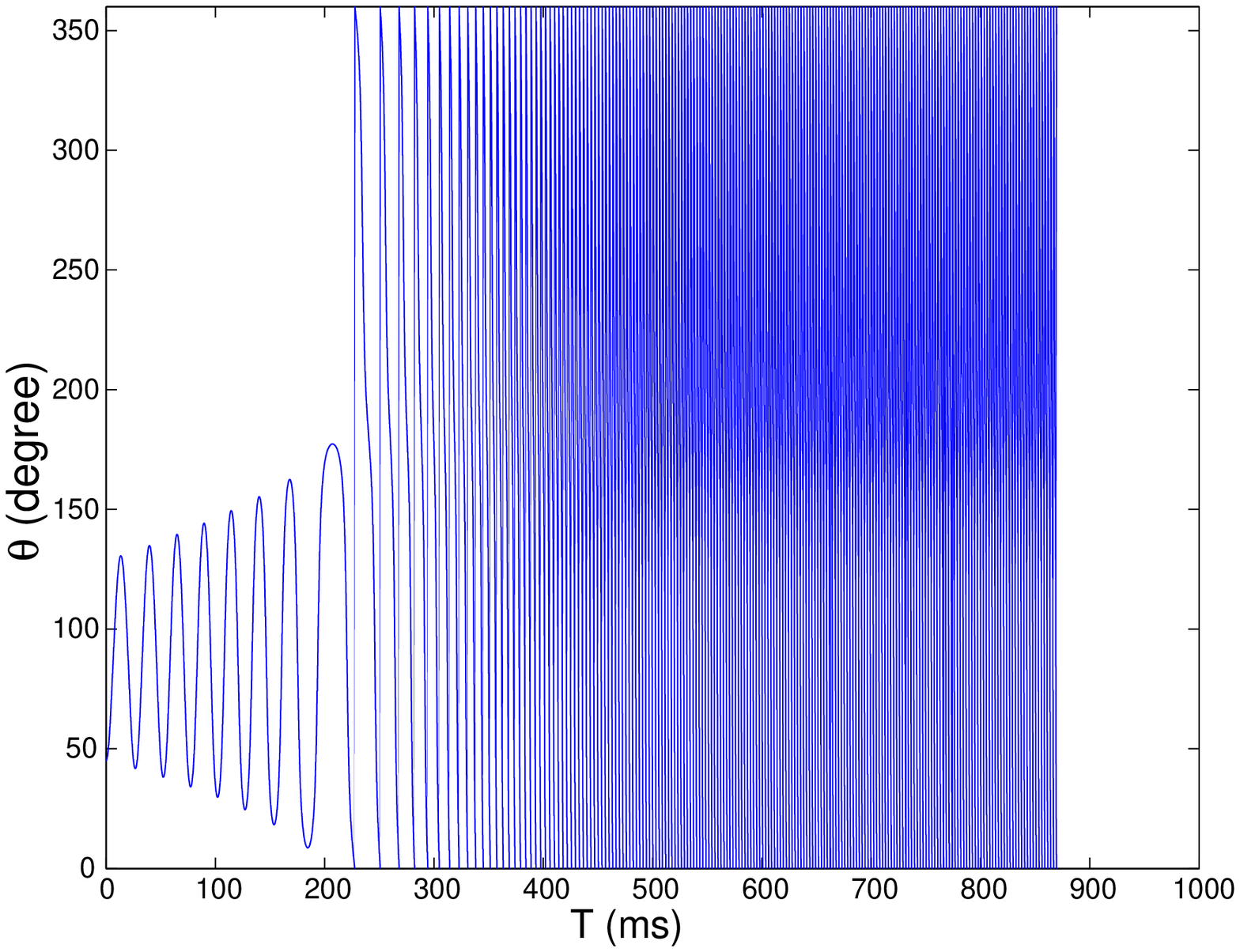}\\
 \end{center}
 \caption{(Color online). History of particle mass center and inclination angle of the biconcave
particle with $s^*=0.481$(top two) and the elliptic particle with $s^*=0.9$ (bottom two) in 
a channel of height 20 $\rm{\mu m}$.}\label{ptk.ctr.y2}
\end{figure}

\begin{figure}
 \begin{center}
    \leavevmode
    \epsfxsize=2.9in \epsffile{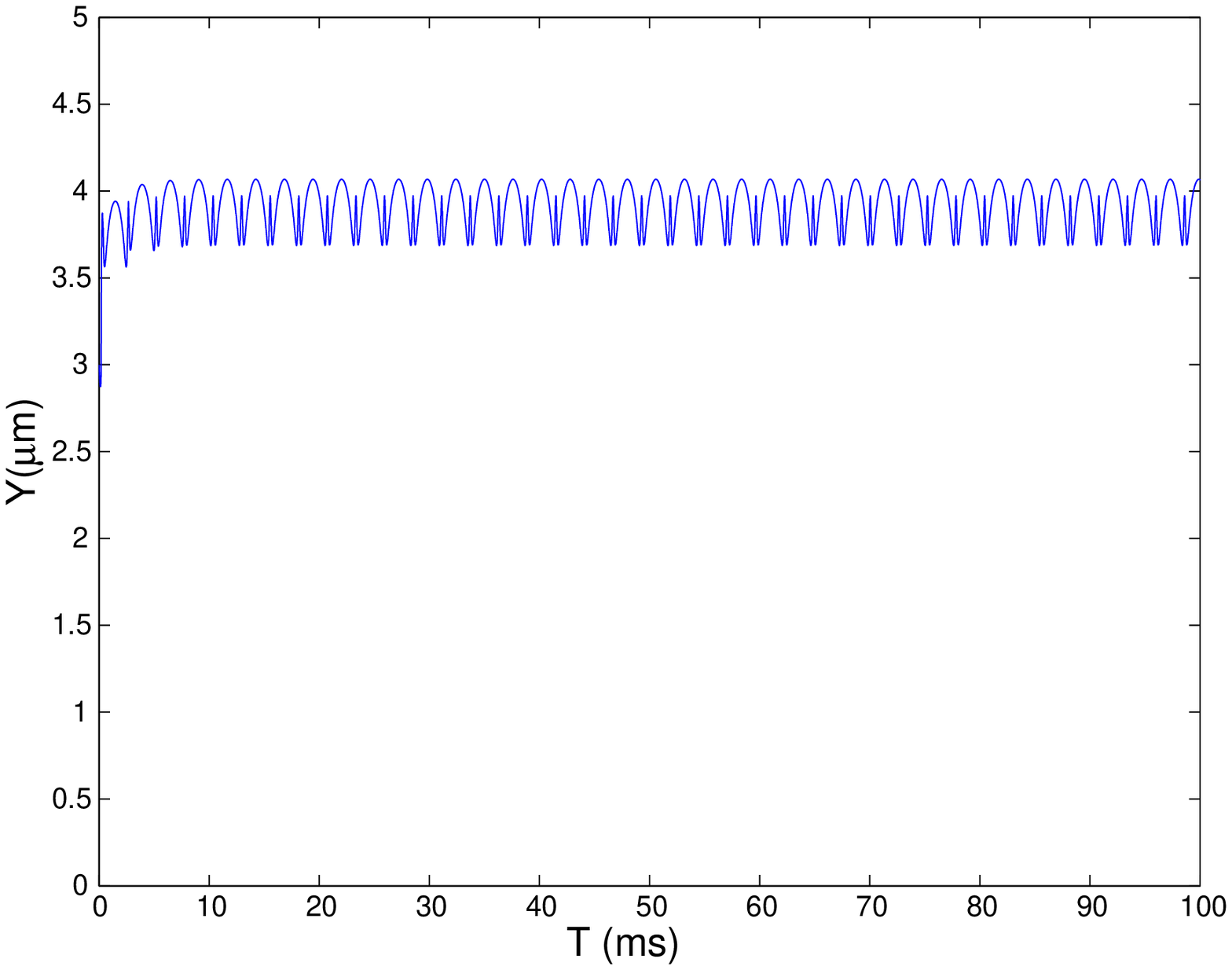}
    \epsfxsize=2.9in \epsffile{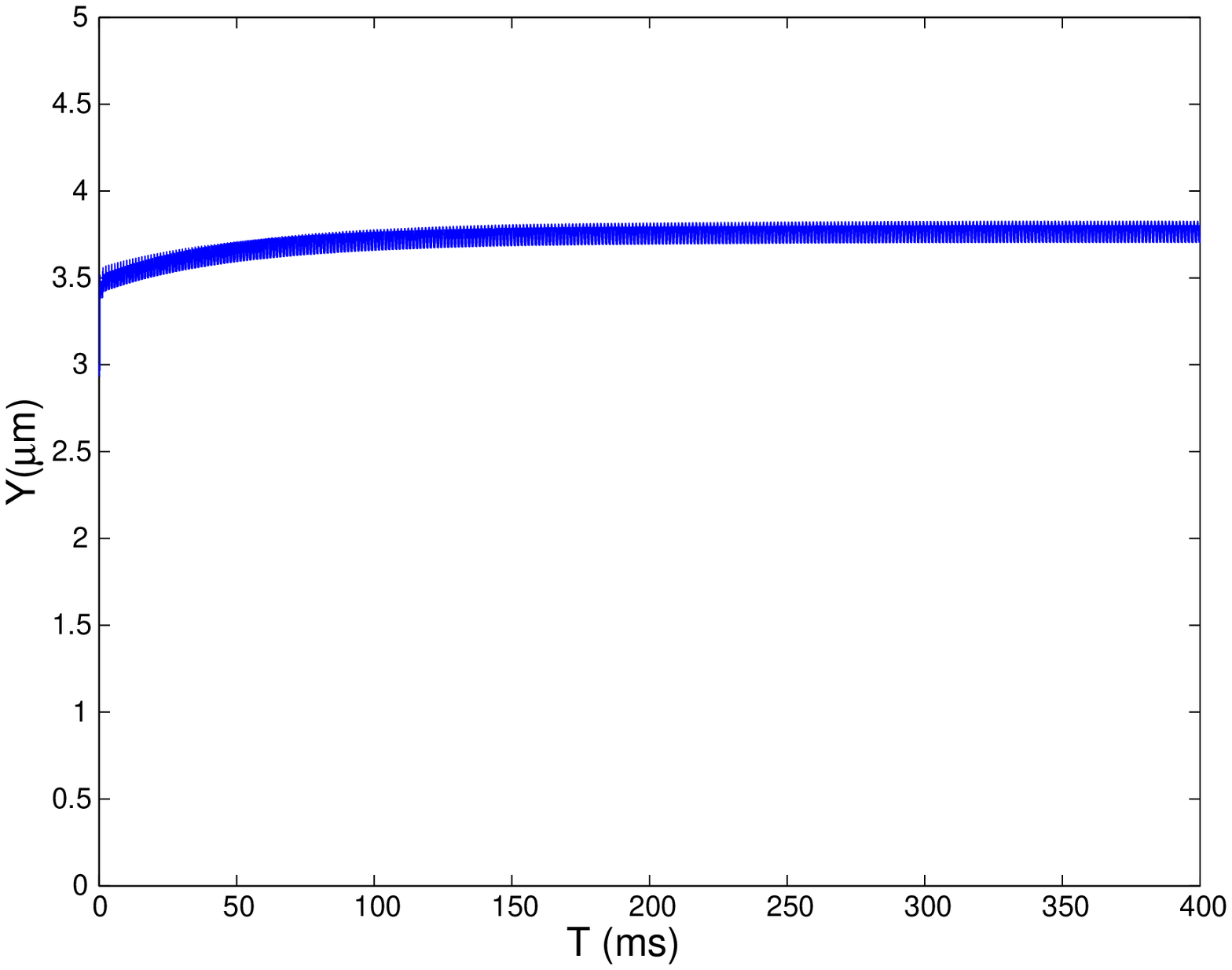}\\
 \end{center}
 \caption{(Color online). History of particle mass center of the biconcave particle with
$s^*=0.481$ (left) and the elliptic particle with $s^*=0.9$ (right). The initial mass center is off the centerline.}\label{ptk.sid.y1}
\end{figure}

\begin{figure}
 \begin{center}
    \leavevmode
     \epsfxsize=5in \epsffile{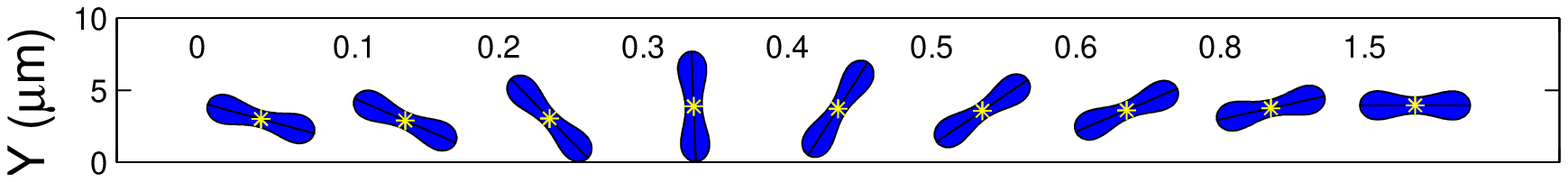}\\
     \epsfxsize=5in \epsffile{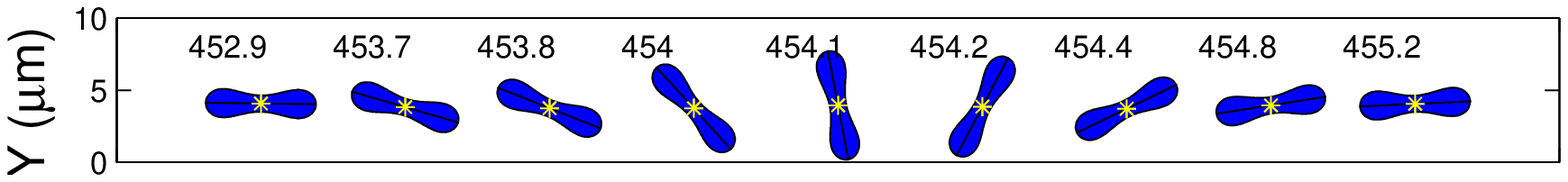}\\
     \epsfxsize=5in \epsffile{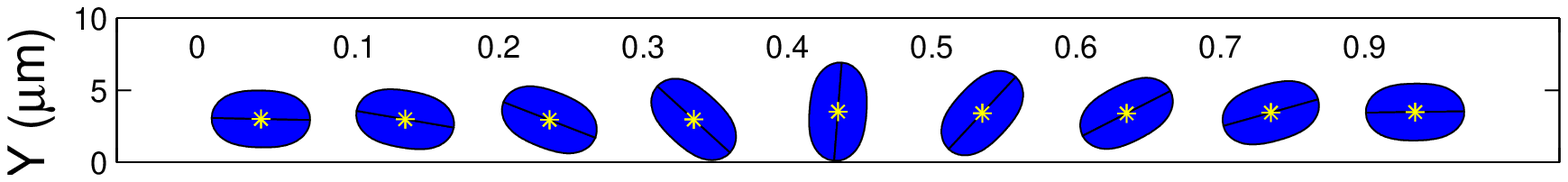}\\
     \epsfxsize=5in \epsffile{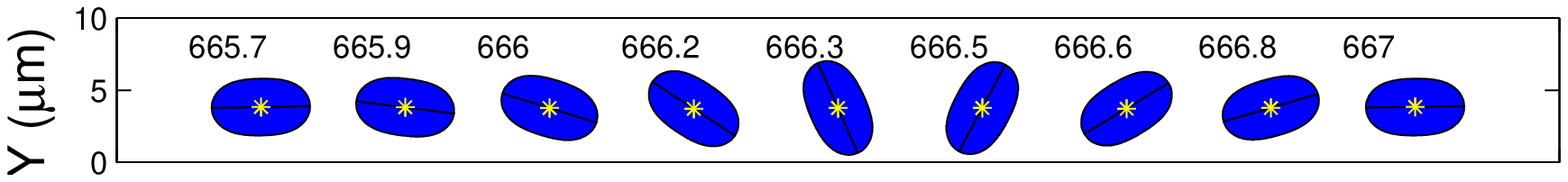}\\ 
 \end{center}
 \caption{(Color online). Snapshots of particle motion in narrow channel with the initial position of the mass center away from
the channel centerline.  The number above each snapshot denotes when the snapshot
was taken(time unit is ms).}\label{snapshot.sid.y1}
\end{figure}

\begin{figure}
 \begin{center}
    \leavevmode
    \epsfxsize=2.9in \epsffile{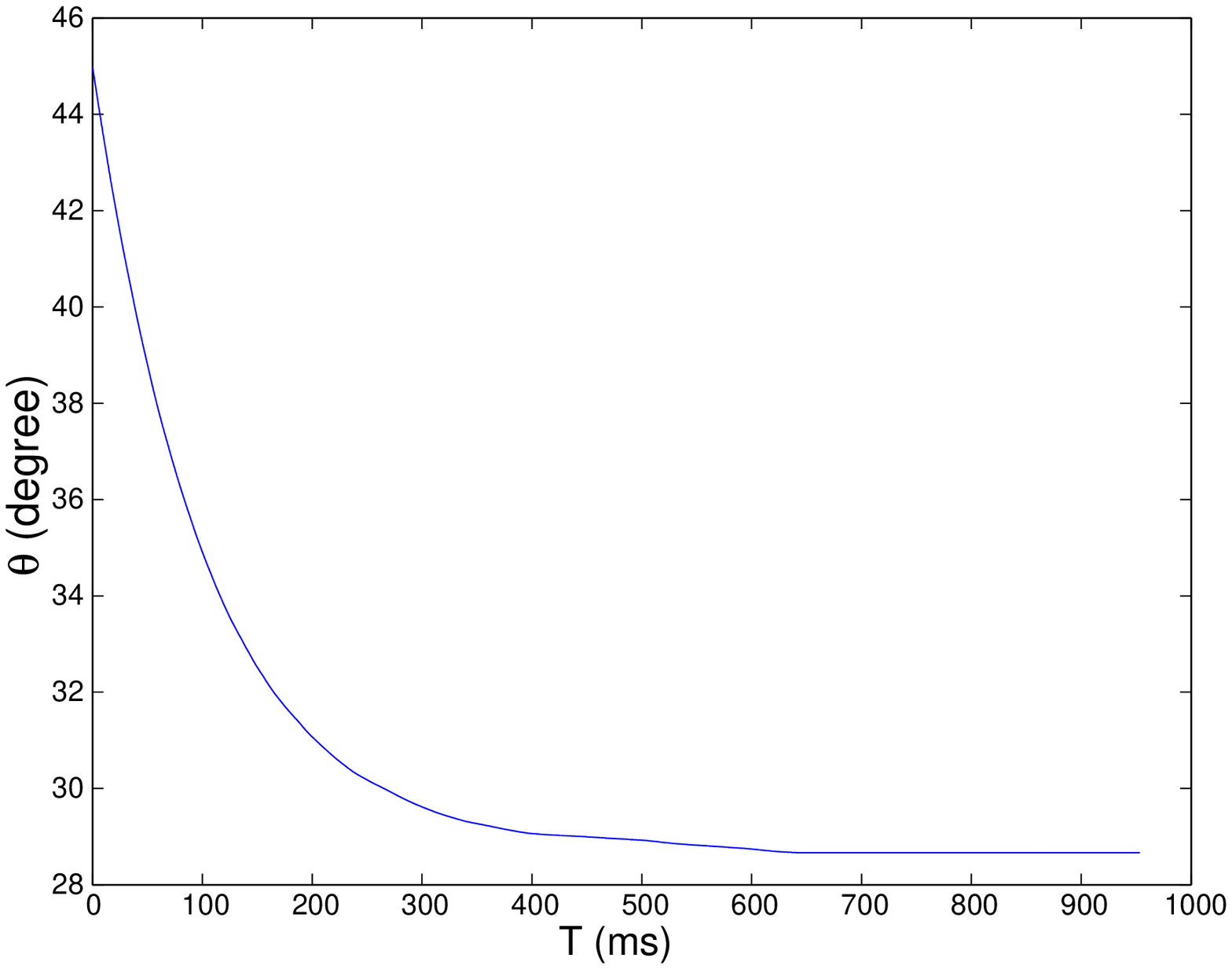}
    \epsfxsize=2.9in \epsffile{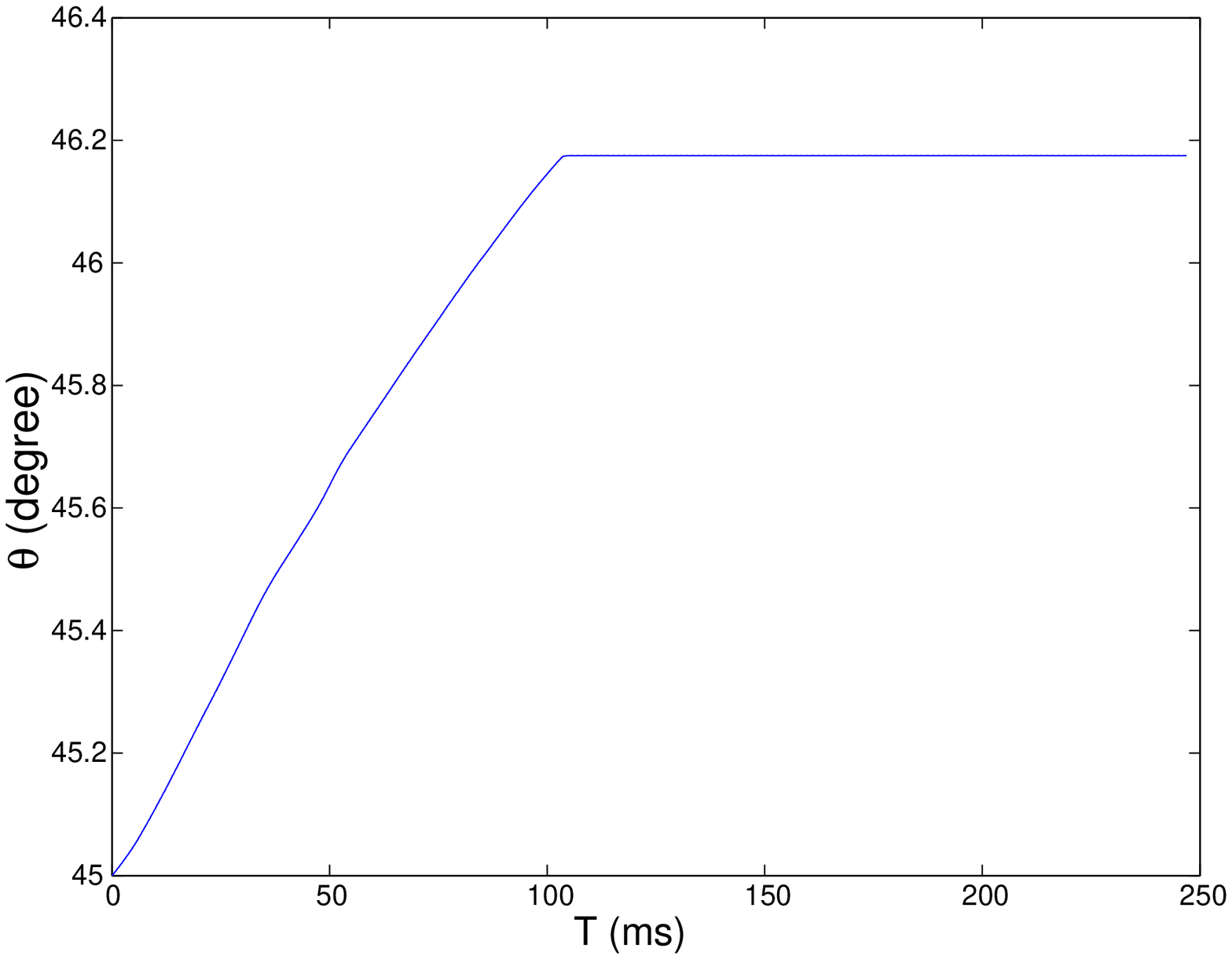}\\
 \end{center}
 \caption{(Color online). History of particle inclination angle when forced to stay at channel center with
$s^*=0.481$ (left)
 and $s^*=0.9$ (right)}\label{fix.ang.y1}
\end{figure}

%%%%%

The motion of a neutrally buoyant particle of either biconcave or elliptical shape in a narrow channel
(100 $\times$ 10 $\rm{\mu m}^2$) has been studied in this section.  The biconcave and elliptical 
shapes are obtained  based on the cell resting shapes used in the previous section.
The long axis of the biconcave shape (resp., elliptical shape) is 7.65 $\rm{\mu m}$ 
(resp., 6.825$\rm{\mu m}$). The particle is suspended in an incompressible Newtonian fluid of 
the density $\rho=1.00$ $\rm{g/cm^{3}}$  and the dynamical viscosity  $\mu$ = 0.012 $\rm{g/(cm s)}$. 
A constant pressure gradient is chosen as in the previous section so that the $Re$ of the flow without 
particle is about 0.4167. 

We have first considered the motions of the particle mass center of both shapes initially 
located at the centerline of the channel with the initial inclination angle $\theta=\pi/4$.
As in Figures \ref{ptk.ctr.y1} and \ref{snapshot.ctr.y1},
both particles show an oscillating motion of the mass center about the
channel center. The up and down motion together with the Poiseuille flow velocity profile
cause the particles to oscillate in the inclination angle.
As the up and down motion becomes stronger, the oscillation in inclination
angle keeps increasing its range. For the particle of biconcave shape, the oscillation 
of inclination angle is first between 0 and 90 degrees and then the particle starts to 
swing back and forth with the inclination angle oscillating between 0 and 180 degrees. 
Once the particle turns horizontal, it starts a tumbling motion and ends 
the oscillation of the  inclination angle. The mass center of the tumbling particle migrates 
toward an equilibrium 
position between the channel centerline and the wall. The initial inclination angle $\theta=\pi/4$
helps the fluid flow to create the oscillation of the particle mass center as shown in Figure  \ref{ang.vs.ycr}.
At the beginning once the mass center is pushing down by the Poiseuille flow, the inclination angle 
is decreasing since the upper portion of the particle close to the centerline moves forward faster than the lower 
portion away from the centerline does. The wall effect from the bottom wall slows down the lateral migration  and finally 
push the neutrally buoyant particle away. When it crosses the centerline, the angular velocity
changes sign accordingly as in Figure \ref{ang.vs.ycr}. 
The neutrally buoyant particle goes through the same up and down motion with rotation changing its 
direction accordingly which is similar to the one in the Stokes regime in \cite{Sugihara-Seki1993}.
But the neutrally buoyant particle moves up and down about the centerline with increasing amplitude 
and finally breaks away from the centerline and moves to  its equilibrium position
between the centerline and the wall as shown in Figure \ref{ptk.ctr.y1}. 
The elliptic shape particle has a similar motion, 
but the oscillation of the inclination angle   is always a back and forth swing motion before the tumbling 
motion occurs. This difference in motion is due to the shape difference of the particles. The particle of 
biconcave shape undertakes stronger resistance force from the flow while the flow can smoothly bypass 
the  particle of elliptic shape. This also explains why it takes the  particle of 
biconcave shape longer time to turn to a tumbling motion than the particle of elliptic shape does.

When increasing the channel height to 20 $\rm{\mu m}$  and 
keeping the other parameters same,  both the up and down motion of the mass center  and the rotation of the long body
oscillates stronger since the wall effect is weaker in a twice wider channel. The neutrally buoyant particle
of long body shape behaves similarly but they migrate away from the centerline faster than they do in the channel of 
height 10 $\rm{\mu m}$.

When the initial mass center is off the centerline, the particle directly turns into
a tumbling motion and migrates toward the equilibrium position between the channel centerline
and the wall in Figure \ref{ptk.sid.y1}. When the particle reaches its equilibrium position, it shows a rotation with 
periodically varying angular velocity as in \cite{Chen2012}. The neutrally buoyant particle behavior 
is entirely different from the cell motion discussed in the previous section due to the lack of the deformability. 
In Figure \ref{snapshot.sid.y1}, two plots of snapshots of the tumbling motion are
shown for each swelling ratio. The first plot is for the first tumbling motion
of the particle where as the second plot is the snapshot of one tumbling motion
when the particle has reached its equilibrium position. By comparing the time in 
the first and second plots, one can observe that it takes longer time for the particle 
to spin a whole circle at the equilibrium position than in the first few spins. This is 
because under the parabolic velocity profile, the flow shear rate is higher when the 
particle is further away from the channel center. So the angular velocity is larger in 
the first few spins since the position of the particle is closer to the channel wall.  

To further study the effect of the up and down motion about the centerline of the mass center, we have added a constraint
on the particle motion so that the particle is only allowed to move freely in the horizontal 
direction and rotate freely. In Figure \ref{fix.ang.y1}, the inclination angle of the constrained motion of
a neutrally buoyant particle reaches to an equilibrium angle without any oscillation when moving
along the centerline in the channel. It shows that the up and down motion of the mass 
center in the central region of the Poiseuille flow causes the oscillation of the inclination angle.

\section{Conclusion}
We have compared the oscillating motions of a neutrally buoyant particle 
and a red blood cell in Poiseuille flow  in a narrow channel
to understand the oscillating motions in \cite{Shi2012a,Shi2012b} and to find out 
the difference between the cell motion and the particle motion.
For the motion of a neutrally buoyant particle of either biconcave or elliptical shape, we 
have obtained  oscillating motion in Poiseuille flow
when the particle mass center is placed at the centerline initially. 
But the neutrally buoyant particle moves up and down about the centerline with the amplitude which 
is increasing in time and finally breaks away from the centerline and moves to its equilibrium position
between the centerline and the wall. The neutrally buoyant particle behavior 
is entirely different from the cell motion due to the lack of the deformability. 
Concerning  the cell motion in the central region of the channel, the oscillating motion occurs 
exactly like a long rigid body as long as the cell mass center moves up and down in the channel 
central region and the cell can maintain a long body shape. But when the mass center 
of the neutrally buoyant particle of a long body shape is not allowed to move up and down, its 
inclination angle reaches a fixed angle without any oscillation. Thus the up-and-down motion of the 
mass center in the channel central region triggers the oscillation motion of a long body entity in Poiseuille flows.

\section*{Acknowledgments}
This work is supported by an NSF grant DMS-0914788. We acknowledge the helpful comments of Chien-Cheng Chang,  Shih-Di Chen, James Feng,
Ming-Chih Lai and Sheldon X. Wang.

\end{document}